\documentclass[aps,prb,twocolumn,floatfix,groupedaddress,longbibliography]{revtex4-2}

\usepackage{amssymb}
\usepackage{amsmath}
\usepackage{bm}
\usepackage{graphicx}
\usepackage{dcolumn}
\usepackage{tikz}
\usepackage[breaklinks=true,bookmarks=true,colorlinks=true,linkcolor=blue,citecolor=red,urlcolor=blue]{hyperref}
\usepackage{xcolor}
\usepackage{physics}
\usepackage{siunitx}

\allowdisplaybreaks % pagebreak aligned equations

\renewcommand\vec{\mathbf} % Vector for alphabets
\newcommand{\uvec}[1]{\hat{\mathbf{#1}}} % Unit vecrors
\newcommand{\hc}{\text{H.c.}}

\usepackage[normalem]{ulem}

\begin{document}

% Title of the article 
%\title{Two-dimensional antiferromagnet with an exact staggered dimer ground state}
\title{Exact columnar dimer ground state and quantum phase transitions in a frustrated coupled spin ladder model}
% Coupled Spin Ladders: Exact Dimer Model and Quantum Phase Transitions
% Exact Dimer Ground State and Phase Transitions in a Frustrated Spin Ladder Model
% Quantum phase transitions in a frustrated coupled-ladder magnet with exact dimer ground state
\author{Manas Ranjan Mahapatra}
\email[]{2020phdph007@curaj.ac.in}
\author{Rakesh Kumar}
\affiliation{School of Physical Sciences, Central University of Rajasthan, Ajmer 305817, India}

\date{\today}

\begin{abstract}
We study a spin-half frustrated coupled ladder system, in which ladders with leg, rung, and diagonal interactions are linked via nearest-neighbor coupling. By introducing a leg-symmetric inter-ladder interaction that connects the left-to-left and right-to-right legs of adjacent ladders, the model is found to possess an exact dimer ground state, characterized by a product of two-spin singlets forming a columnar dimer phase. We analyze this model using bond-operator mean-field theory (BOMFT) and the density matrix renormalization group (DMRG) to probe the phase transitions that occur as one traverses the coupling space. The BOMFT reveals three distinct phases: a double-stripe ordered phase, a Néel ordered phase, and a quantum disordered dimerized phase. The critical points for the transitions are at \( J_1 = -0.81 \) (double-stripe to dimerized) and at \( J_1 = 2.81 \) (dimerized to Néel phase). Further, the DMRG results corroborate the exact ground state and refine the critical points to \( J_1 = -0.79 \) and \( J_1 = 2.29 \) for the respective transitions. Additionally, another transition is identified as the Néel order vanishes for \( J_1 \ge 4.5 \). The model can alternatively be represented as a network of orthogonal zigzag and fully frustrated spin ladders, offering a structural framework conducive to quantum materials engineering.
\end{abstract}


% maketitle must follow title, authors, abstract, \pacs, and \keywords
\maketitle

% Main text: Sections, subsections, subsubsections
\section{Introduction} % (fold)
\label{sec:introduction}

The exact solution for a one-dimensional spin-half Heisenberg antiferromagnet, as given by Bethe, reveals that there is no true long-range order due to quantum fluctuations. Instead, the spin-spin correlation decays algebraically with the distance between spins~\cite{bethe1931theory}. When coupling multiple chains to form a spin ladder, the two-leg spin ladder system becomes gapped, meaning a finite energy is required to create an $S=1$ excitation. In cases where the rung interactions $J'$ are stronger than the chain interactions $J$, the ground state is a product of spin singlets on the rungs, with a total spin $S=0$. Breaking a rung singlet generates an $S=1$ triplet excitation~\cite{PhysRevB.45.5744}. It was predicted that the spin gap vanishes only when $J'=0$, and for any $J'>0$, the system remains gapped~\cite{PhysRevB.47.3196}. Unlike spin chains, spin ladders exhibit purely short-range order, with spin-spin correlations decaying exponentially. This result has been confirmed through numerous numerical techniques and experimentally observed in compounds such as SrCu$_2$O$_3$~\cite{Azuma1994}, (VO)$_2$P$_2$O$_7$~\cite{PhysRevB.35.219}, and LaCuO$_{2.5}$~\cite{PhysRevB.55.R6117}. While these studies are focused on spin ladders without frustrated interactions, more recent research has explored the antiferromagnetic Heisenberg model in spin ladders with frustration, such as in the compound BaFe$_2$Se$_3$, where diagonal interactions, along with next-nearest-neighbor interactions along the leg, are present~\cite{PhysRevB.108.014416}. Additionally, BiCu$_2$PO$_6$ has been studied using a Hamiltonian that includes Heisenberg interactions along with Dzyaloshinskii-Moriya (DM) and anisotropic superexchange interactions. In this system, frustration arises from second-nearest-neighbor chain interactions~\cite{Tsirlin2010,pikulski2020two}.

Frustrated $S=\frac{1}{2}$ ladder systems have been studied extensively as platforms for investigating the interplay between competing interactions and quantum fluctuations in low-dimensional magnets. Early work by Gelfand~\cite{Gelfand1991}, building on the foundational notions of Shastry and Sutherland ~\cite{shastry1981}, demonstrated that when a Hamiltonian can be decomposed into elementary triangular units, the frustration arising from leg and diagonal couplings can stabilize exact dimerized ground states at specific points in the parameter space. This established one of the first analytically tractable examples of frustrated ladders.

Subsequent studies have explored the rich physics of such systems under varying conditions, particularly in the presence of an external magnetic field. In appropriate limits, both frustrated and unfrustrated ladders can be mapped onto effective XXZ models, providing a unified framework for understanding magnetization processes and plateau formation~\cite{Mila1998}. These systems also exhibit a variety of unconventional magnetic responses have been identified, including magnetization plateaux, jumps, and complex excitation spectra involving magnons and spinons~\cite{Honecker2000,Fouet2006,Michaud2010}. Frustration further stabilizes fractional plateaux over extended parameter regimes and can give rise to phenomena such as quantum bicriticality in strongly frustrated systems~\cite{Almeida2023}.

In the absence of a magnetic field, competing interactions in ladder systems lead to a rich interplay between magnetic order and dimerization. Frustration introduced via diagonal interchain couplings generates competing magnetic and dimerized phases that extend beyond the conventional rung-singlet and Haldane regimes~\cite{li2012quantum,hikihara2010phase}. Specifically, diagonal frustration can induce staggered dimer order, though its stability is typically restricted to narrow regions of the phase diagram. The effects of frustration are further enriched when additional interactions, such as next-nearest-neighbor couplings along the legs are included. Studies of these extended ladder models have demonstrated the emergence of both columnar and staggered dimer phases~\cite{PhysRevB.73.214427,PhysRevB.86.075133,liu2008}. More recently, advances in numerical techniques—including quantum Monte Carlo simulations in specialized bases—have enabled the detailed exploration of highly frustrated ladders, revealing complex phase diagrams where rung-singlet, rung-triplet, and magnetically ordered phases compete~\cite{wessel2017}. These developments emphasize the diversity of ground states in frustrated ladders and their extreme sensitivity to the precise form of competing interactions.

Beyond purely theoretical investigations, experimental realizations of frustrated and weakly coupled ladder systems have further highlighted the richness of ladder physics. Recent NMR measurements on weakly coupled ladders revealed an unexpected crossover within the ordered phase, demonstrating how anisotropic and frustrated inter-ladder couplings can significantly modify low-temperature magnetic correlations~\cite{jeong2017magnetic}. Similarly, inelastic neutron scattering experiments on Ba$_2$CuTeO$_6$ identified a quantum critical point separating a gapped ladder regime from a long-range Néel-ordered phase, providing direct evidence that inter-ladder coupling can drive dimensional crossover and magnetic ordering~\cite{macdougal2018spin}. These experimental findings emphasize the need for controlled theoretical models that systematically incorporate frustration, anisotropy, and inter-ladder coupling in order to clarify the mechanisms governing quantum phase transitions in ladder-based systems.

Motivated by theoretical studies of frustrated ladder systems and experimental observations of coupled-ladder compounds, which highlight the critical roles of inter-ladder coupling and dimensional crossover, we propose a spin-$\frac{1}{2}$ coupled-ladder Heisenberg antiferromagnet. In this model, ladders with leg, rung, and diagonal interactions are further linked via a horizontal inter-ladder coupling. By introducing a spatially anisotropic third-nearest-neighbor interaction along the horizontal direction, we construct a model that admits an exact columnar dimer ground state, characterized by a direct product of singlet pairs on the rungs. In contrast to previously studied models where dimerized phases often occupy narrow parameter regimes, our construction provides a robust realization of a columnar dimer phase in a coupled-ladder geometry. This allows a systematic investigation of its stability against competing magnetic orders and provides a controlled platform for studying quantum phase transitions between dimerized and magnetically ordered states.

The remainder of this paper is structured as follows. The model and its ground state with supporting exact diagonalization data are placed in Sec.~\ref{sec:model}. After that, the bond-operator mean-field calculations are presented in Sec.~\ref{sec:bond_operator_mean_field_theory}. Subsequently, the mean-field and DMRG results with analyses are provided in Sec.~\ref{sec:results_and_discussion}. Finally, we conclude this work in Sec.~\ref{sec:conclusions}.

\section{Model}
\label{sec:model}

\begin{figure}[h]
  \centering
    \includegraphics[width=0.4\textwidth]{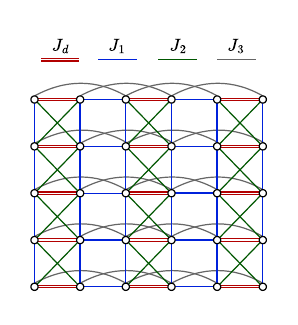}
  \caption{Schematic representation of the coupled ladder model described by Eq.~\eqref{eq:model_hamiltonian}. The double red lines denote the intra-ladder rung exchange interactions ($J_d$), while the nearest-neighbor leg couplings ($J_1$) are shown in blue. The same coupling $J_1$ also mediates interactions between neighboring legs of adjacent ladders. The green lines represent the diagonal intra-ladder frustrating interactions ($J_2$), and the dark grey lines correspond to the long-range inter-ladder couplings ($J_3$).}
  \label{fig:dimer_lattices}
\end{figure}

The model consists of coupled two-leg spin-$\frac{1}{2}$ Heisenberg ladders, as illustrated in Fig.~\ref{fig:dimer_lattices}. Each ladder contains three types of exchange interactions: the leg interaction $J_1$, the rung interaction $J_d$, and the diagonal intra-ladder interaction $J_2$. Adjacent ladders are further coupled through inter-ladder interactions $J_1$ and $J_3$, where $J_1$ connects neighboring legs of adjacent ladders, while $J_3$ couples alternate legs along the horizontal direction. Notably, the exchange strength along an individual leg and between neighboring legs of adjacent ladders is identical and given by $J_1$.

Alternatively, the system may be viewed as an anisotropic square lattice with nearest-neighbor couplings ($J_d$, $J_1$), next-nearest-neighbor frustrating interactions ($J_2$), and third-nearest-neighbor interactions ($J_3$). The Hamiltonian of the model is given by
\begin{equation}
    H = \sum_{\langle i, j \rangle} J_{ij} \Vec{S}_i \cdot \Vec{S}_j + H^\prime,
    \label{eq:model_hamiltonian}
\end{equation}
where the nearest-neighbor exchange couplings $J_{ij}$ exhibit spatial anisotropy along the horizontal direction, alternating between $J_d$ and $J_1$, while remaining uniform with strength $J_1$ along the vertical direction. The additional interaction term $H^\prime$ consists of two contributions,
\begin{equation}
    H' = J_2 {\sum_{\langle i, j \rangle_2}}^\prime \Vec{S}_i \cdot \Vec{S}_j + J_3 {\sum_{\langle i, j \rangle_3}}^\prime \Vec{S}_i \cdot \Vec{S}_j,
\label{eq:sub_model_hamiltonian}
\end{equation}
where, the first primed sum denotes a summation over a restricted subset of diagonal next-nearest-neighbor pairs, while the second primed sum represents a sum over longitudinal third-nearest neighbors constrained to the horizontal axis.

\begin{figure}[h]
  \centering
    \includegraphics[width=.45\textwidth]{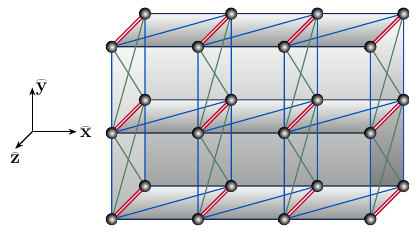}
  \caption{Alternative bilayer representation mapped from the two-dimensional model~\eqref{eq:model_hamiltonian}, where fully frustrated ladders oriented along the $yz$ planes intersect orthogonally with zigzag ladders in the $zx$ planes. The inter-ladder interactions in the 2D model form zigzag ladders in the bilayer representation; specifically, the leg-symmetric third-nearest-neighbor and inter-ladder nearest-neighbor interactions correspond to interactions along the legs and diagonals of the zigzag ladders, respectively. Furthermore, both types of ladders share common rungs. Consequently, the overall mapping preserves the complete topology of the original interactions.}
  \label{fig:Model3D}
\end{figure}

The two-dimensional model Hamiltonian~\eqref{eq:model_hamiltonian} can be physically understood better by mapping it onto an alternative three-dimensional bilayer geometry, as schematically illustrated in Fig.~\ref{fig:Model3D}. In this rearranged lattice configuration, the original 2D network decomposes into two sets of mutually orthogonal, intersecting ladder systems. Specifically, the intra-ladder terms form fully frustrated ladders that lie entirely within the $yz$ planes, while the inter-ladder coupling terms are mapped onto zigzag ladders spanning the $zx$ planes. Within these zigzag structures, the leg-symmetric third-nearest-neighbor and the inter-ladder nearest-neighbor interactions of the original 2D model map directly onto the leg and diagonal bonds, respectively. Crucially, these two orthogonal ladder subsystems intersect such that they share common rungs parallel to the $\uvec{z}$ axis. This geometric arrangement ensures a one-to-one correspondence between the bonds of both representations, thereby preserving the complete topology and connectivity of the original interactions.

As illustrated in Fig.~\ref{fig:Model3D}, the fully frustrated and zigzag ladders contribute $4N_d$ and $2N_d$ triangles, respectively, with localized spins residing on their vertices, where $N_d$ denotes the number of dimer bonds with coupling strength $J_d$. This structural decomposition allows the total Hamiltonian given in Eq.~\eqref{eq:model_hamiltonian} to be partitioned into a sum of cluster Hamiltonians, $h_m$, defined over individual triangles:
\begin{align}
  H&= \sum_{m=1}^{4N_d} h_m^{(yz)}  + \sum_{m=1}^{2N_d} h_m^{(zx)}- J_d\sum_{\langle i,j \rangle_d}^{}\vec{S}_i \cdot\vec{S}_j,
  \label{eq:tri_ham}
\end{align}
where the cluster Hamiltonians for the respective planes are expressed as:
\begin{align}
   h_m^{(yz)} (i,j,k)=\frac{J_d}{4}\vec{S}_{i}\cdot \vec{S}_j+\frac{J_1}{2}\vec{S}_j \cdot\vec{S}_k+\frac{J_2}{2}\vec{S}_{k}\cdot\vec{S}_i,\label{eq:tri_yz}\\
   h_m^{(zx)} (i,j,k)=\frac{J_d}{2}\vec{S}_{i}\cdot \vec{S}_j+J_3\vec{S}_j \cdot\vec{S}_k+\frac{J_1}{2}\vec{S}_{k}\cdot\vec{S}_i.\label{eq:tri_zx}
\end{align}
While each spin-spin interaction enters the original lattice Hamiltonian exactly once, the triangle-based representation introduces structural overlaps. To compensate for this overcounting, appropriate scaling factors are introduced in the denominators of the coupling constants in Eqs.~\eqref{eq:tri_yz} and~\eqref{eq:tri_zx}. Finally, the subtractive last term in Eq.~\eqref{eq:tri_ham} explicitly corrects for the residual double-counting of the dimer bonds, where $\sum_{\langle i,j \rangle_d}$ denotes the summation over all dimer bonds. On simplifying the Hamiltonian, we get the following form at $J_1=J_2=2J_3$:
\begin{align}
  H&=-\frac{9}{4}J_1 N_d+ 3J_1 \left(\frac{J_d}{3J_1} -1\right) \sum_{\langle i,j\rangle_d}^{}\vec{S}_i \cdot\vec{S}_j\nonumber\\
  & + \frac{3}{4}J_1\left[\sum_{\triangle_{yz}^{}}P_{3/2}(\triangle_{yz}) + \sum_{\triangle_{zx}^{}}P_{3/2}(\triangle_{zx})\right],
  \label{eq:hamil_proj}
\end{align}
where the projection operator on a three-spin cluster is
\begin{equation}
  P_{3/2}(\triangle)= \frac{1}{2}+\frac{2}{3}\left[\vec{S}_i \cdot \vec{S}_j + \vec{S}_j \cdot \vec{S}_k+\vec{S}_{k} \cdot\vec{S}_i\right].
  \label{eq:projop}
\end{equation}

\begin{table}
\centering
\begin{ruledtabular}
\begin{tabular}{ll}
\multicolumn{1}{l}{\textrm{Total Spin ($S_{\triangle}$)}} &
\multicolumn{1}{l}{\textrm{Eigenstates}} \\
\hline
$\frac{3}{2}$ (one quartet) & $|\uparrow\uparrow\uparrow\rangle$ \\
          & $|\downarrow\downarrow\downarrow\rangle$ \\
         & $\frac{1}{\sqrt{3}}(|\uparrow\uparrow\downarrow\rangle+|\uparrow\downarrow\uparrow\rangle+|\downarrow\uparrow\uparrow\rangle) $\\
          & $\frac{1}{\sqrt{3}}(|\uparrow\downarrow\downarrow\rangle+|\downarrow\downarrow\uparrow\rangle+|\downarrow\uparrow\downarrow\rangle)$ \\
\hline
$\frac{1}{2}$ (two doublets) & $\frac{1}{\sqrt{2}}(|\uparrow\uparrow\downarrow\rangle-|\downarrow\uparrow\uparrow\rangle)$ \\
          & ${\frac{1}{\sqrt{2}}(|\downarrow\uparrow\uparrow\rangle - |\uparrow\uparrow\downarrow\rangle)}$ \\
          & ${\sqrt{\frac{2}{3}}|\downarrow\uparrow\downarrow\rangle-\frac{1}{\sqrt{6}}(|\uparrow\downarrow\downarrow\rangle + |\downarrow\downarrow\uparrow\rangle)}$ \\
          & ${\sqrt{\frac{2}{3}}|\uparrow\downarrow\uparrow\rangle-\frac{1}{\sqrt{6}}(|\uparrow\uparrow\downarrow\rangle + |\downarrow\uparrow\uparrow\rangle)}$ \\
\end{tabular}
\end{ruledtabular}
\caption{Eigenstates and their corresponding total spins for a cluster of three $S=\frac{1}{2} $ spins.}
\label{tab:spin_eigenstates}
\end{table}

A cluster of three $S=1/2$ spins spans an 8-dimensional Hilbert space that can be partitioned into doublet and quartet subspaces (see Table~\ref{tab:spin_eigenstates}), specifically consisting of two doublets ($S_{\triangle}=1/2$) and one quartet ($S_{\triangle}=3/2$). The projection operator $P_{3/2}(\triangle)$ defined in Eq.~\eqref{eq:projop} projects directly onto this quartet subspace and annihilates any doublet state. Among the four total doublet states, the energy is minimized exclusively by those configurations where a specific pair of spins forms a spin singlet while the remaining spin remains free~\cite{PhysRevB.110.104402}. This free spin on one triangle can subsequently form a singlet with a corresponding free spin from an adjacent triangle. Consequently, minimizing the total energy requires finding states that yield the lowest possible expectation value of the projector, namely $\langle P_{3/2} \rangle = 0$, given the bounded nature of the operator ($0 \leq \langle P_{3/2} \rangle \leq 1$). Let us consider a trial wavefunction constructed as a product of independent spin singlets:
\begin{equation}
\ket{\Psi} = \bigotimes_{\langle i,j \rangle_d} [i,j],
\label{eq:exact_ground_state}
\end{equation}
where $[i,j] \equiv \frac{1}{\sqrt{2}} \left( \ket{\uparrow_i \downarrow_j} - \ket{\downarrow_i \uparrow_j} \right)$ denotes the standard valence-bond spin-singlet state residing on the dimer bonds (or rungs). Because every triangle in the lattice contains at least one such dimer singlet, applying $P_{3/2}$ to $\ket{\Psi}$ yields zero identically. As a result, the projection terms in the total Hamiltonian~\eqref{eq:hamil_proj} do not contribute to the expectation value. Furthermore, $\ket{\Psi}$ is an eigenstate of the remaining dimer summation term, which maximally lowers the energy for all $J_d \geq 3J_1$. Therefore, $\ket{\Psi}$ constitutes the exact ground state of the full Hamiltonian at the highly symmetric point $J_1 = J_2 = 2J_3$ for $J_d \geq 3J_1$ (the Shastry-Sutherland line), yielding a ground-state energy of $E_{\text{gs}} = -\frac{3}{4}J_d N_d$.

The systematic construction of frustrated spin Hamiltonians hosting exact dimer ground states traces back to the pioneering work of Majumdar and Ghosh (MG) for spin-$1/2$ chains~\cite{1969JMP....10.1388M}. This framework was subsequently extended to two dimensions in the landmark Shastry-Sutherland (SS) model~\cite{shastry1981}, which expresses the total Hamiltonian as a sum of interacting three-spin clusters. Over the decades, generalized mathematical frameworks using the representation theory of symmetric groups and spin projection operators have been developed to systematically construct such models~\cite {klein1982exact, chayes1989valence, lowdin1964angular}. This approach continues to inspire extensive analytical, numerical, and experimental investigations into exactly solvable dimerized phases across various spatial dimensions~\cite{takano1994construction, PhysRevB.66.024406, PhysRevB.110.104402}

\begin{figure}[h]
  \centering
    \includegraphics[width=.45\textwidth]{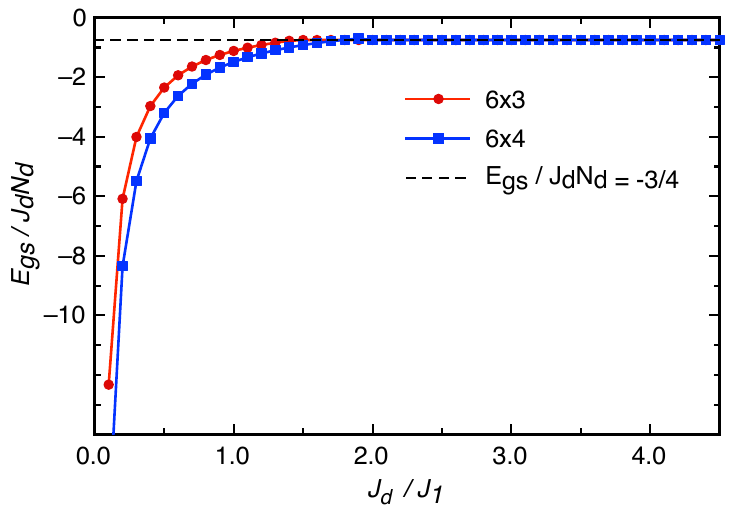}
    \includegraphics[width=.45\textwidth]{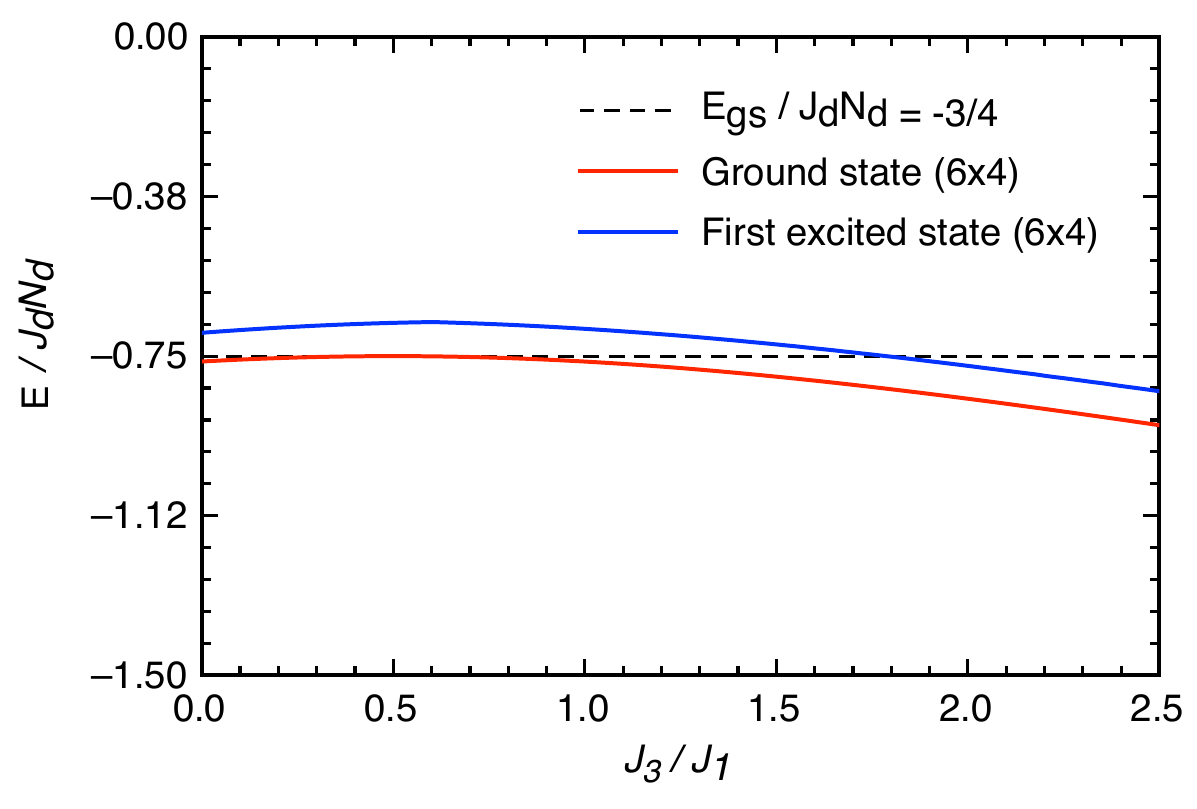}
  \caption{Numerical results obtained via exact diagonalization for $6\times3$ and $6\times4$ clusters. The upper panel displays the normalized ground-state energy per dimer, $E_{\text{gs}} / J_d N_d$, as a function of $J_d / J_1$ at the symmetric point $J_1 = J_2 = 2J_3$, where the analytical value $E_{\text{gs}} / J_d N_d = -3/4$ is denoted by the dashed line; deviations at lower coupling ratios highlight significant finite-size effects. The lower panel illustrates the low-energy spectrum ($E / J_d N_d$) of the $6\times4$ cluster as a function of $J_3 / J_1$ at the fixed threshold $J_d = 3J_1$ (with $J_1 = J_2$). Here, the clear finite separation between the ground state (red line) and the first excited state (blue line) explicitly demonstrates the gapped nature of the exact dimerized phase.}
  \label{fig:edres}
\end{figure}

To validate our analytical findings, we performed exact diagonalization (ED) calculations on $6 \times 3$ and $6 \times 4$ plannar lattices. The numerical results agree well with our analytical predictions (see Fig.~\ref{fig:edres}). Notably, the phase boundary corresponding to the exact dimer ground state extends slightly below the analytical critical threshold of $J_d = 3J_1$ in the ED spectra. We attribute this broadening of the phase boundary to finite-size effects, a conclusion supported by the systematic convergence observed between the two cluster sizes illustrated in Fig.~\ref{fig:edres}. The first excited state is separated from the ground state by a finite energy gap, which is consistent with the expected behaviors of a dimerized phase.

The construction and exploration of quantum ground states, alongside their mutual phase transitions in spin-ladder geometries, have a long and rich history (see Reviews~\cite{Mikeska2004, Miyahara2011}). The conventional, unfrustrated spin ladder featuring only nearest-neighbor antiferromagnetic leg and rung interactions is a prototypically gapped system for any non-zero rung coupling~\cite{Barnes1993, Nishiyama1995, Greven1996}. Correspondingly, an isolated zigzag ladder can be viewed as a generalization of the MG model, where the exchange interactions undergo spatial modulation along the chain direction~\cite{White1996, Brehmer1996, Watanabe1999}. In the zigzag ladders considered here, $J_d$ and $J_1$ alternate along the chain, while $J_3$ serves as the next-nearest-neighbor interaction. Crucially, a zigzag ladder reduces precisely to the standard MG chain when setting $J_1 = J_d = 2J_3$, which hosts a twofold degenerate ground state. Upon introducing bond modulation, \emph{only} one of these two MG configurations survives as an exact ground state, provided that $J_1 = 2J_{3}$ for $J_d > 2J_3$ (or conversely, $J_{d} = 2J_3$ for $J_1 > 2J_3$). These highly symmetric parameter trajectories are known as the SS lines. Furthermore, it has been established that a fully frustrated ladder with symmetric leg couplings ($J_1 = J_2$) exhibits a direct, first-order phase transition separating a rung-triplet phase at weak coupling from a rung-singlet phase at strong coupling, with a critical threshold of $J_d / J_1 \simeq 1.4$~\cite{Gelfand1991, Bose1993, Xian1995, Honecker2000,Honecker2016,wessel2017}. Our current model enriches these paradigms by intertwining the distinct physical behaviors of both fully frustrated and zigzag ladder subsystems.

Having analyzed the exact solvable point and the Shastry-Sutherland lines of the Hamiltonian~\eqref{eq:model_hamiltonian}, we now examine its behavior in various limiting regimes:
\begin{itemize}
  \item Isolated Rung Limit:
   ($J_d \neq 0$, all other $J_i = 0$): The Hamiltonian reduces completely to independent rungs, yielding the unique and exact product state $\ket{\Psi}$ as its ground state.
  \item Decoupled Chain Limits:
\begin{itemize}
  \item If $J_2 \neq 0$ and $J_d = J_1 = J_3 = 0$, the lattice decomposes into non-interacting, uniform spin-1/2 Heisenberg chains, yielding a gapless excitation spectrum.
  \item Similarly, if $J_3 \neq 0$ and $J_d = J_1 = J_2 = 0$, the system again reduces to decoupled spin chains characterized by gapless magnetic excitations.
\end{itemize}
  \item Vanishing Frustration Limit: 
  \begin{itemize}
    \item Setting $J_1 \neq 0$, $J_d = J_2 = J_3 = 0$), the system decouples into independent, unfrustrated two-leg ladders with uniform exchange couplings on both rungs and legs. This regime famously hosts a robust rung-singlet ground state protected by a finite spin gap for any $J_1 > 0$.
    \item Setting only $J_1 = 0$ removes the source of frustration, transforming the system into an interconnected network of anisotropic, unfrustrated two-leg ladders of both types. Due to the lack of frustration and the prevailing ladder geometries, a fully gapped excitation spectrum is expected to persist across this entire regime.
    \item Setting only $J_2 = 0$ removes the specific frustration channel associated with the $J_2$ bonds, resulting in an interconnected network of anisotropic, unfrustrated ladders intertwined with anisotropic zigzag ladders. Due to the reduced frustration and the underlying ladder geometries, the system is expected to yield a fully gapped excitation spectrum.
  \end{itemize}
\end{itemize}

In the rest of the paper, we fix $J_{d}=3$, $J_{2}=1$, and $J_{3}=0.5$, for which the exact dimer condition is satisfied at $J_{1}=1$. We then vary $J_{1}$ around this point, exploring both ferromagnetic and antiferromagnetic regimes in the remainder of the paper. In this way, the exact dimer point serves as the reference configuration, and deviations in $J_1$ probe the robustness of the dimer phase against competing magnetic orders.

\section{Bond-operator mean-field theory} % (fold)
\label{sec:bond_operator_mean_field_theory}
We analyze this model using a low-energy bosonic mean-field theory, focusing on triplet fluctuations around a non-magnetic, dimerized quantum reference state. In our case, the reference state is a columnar dimer on a square lattice. This approach offers a straightforward way to investigate the stability of the reference state against low-energy quantum fluctuations. For a pair of spin-$\frac{1}{2}$ particles, the Hilbert space consists of one singlet and three triplet states. Sachdev and Bhatt introduced bond operators that create these four states, $|s\rangle$, $|t_x\rangle$, $|t_y\rangle$, and $|t_z\rangle$~\cite{PhysRevB.41.9323}, which are given by
\begin{subequations}
  \label{eq:mapping}
  \begin{align}
    s^\dagger\ket{0}=\ket{s}&\equiv \frac{1}{\sqrt{2}}(\ket{\uparrow \downarrow}-\ket{\downarrow \uparrow}),\label{eq:sdef}\\
     t_x^\dagger\ket{0}= \ket{t_x}&\equiv\frac{-1}{\sqrt{2}}(\ket{ \uparrow \uparrow}-\ket{\downarrow \downarrow}),\label{eq:txdef}\\
    t_y^\dagger\ket{0}=  \ket{t_y}&\equiv \frac{i}{\sqrt{2}}(\ket{ \uparrow \uparrow}+\ket{\downarrow \downarrow}),\label{eq:tydef}\\
     t_z^\dagger\ket{0}=\ket{t_z}&\equiv\frac{1}{\sqrt{2}}(\ket{\uparrow \downarrow}+\ket{\downarrow \uparrow}).\label{eq:tzdef}
  \end{align}
\end{subequations}
 These operators obey bosonic commutation relations. Using this formalism, the spin operators are expressed as:
\begin{equation}
  S_{1\alpha}=\frac{1}{2}(s^\dag t_\alpha+t_\alpha^ \dag s-i\epsilon_{\alpha\beta\gamma} t_\beta ^\dag t_\gamma)
  \label{eq:bo_representation_1}
\end{equation}
\begin{equation}
  S_{2\alpha}=\frac{1}{2}(-s^\dag t_\alpha-t_\alpha ^\dag s-i\epsilon_{\alpha\beta\gamma} t_\beta ^\dag t_\gamma)
  \label{eq:bo_representation_2}
\end{equation}
where $\alpha,\beta,\gamma\in\{x,y,z\}$, $\epsilon_{\alpha\beta\gamma}$ is the totally antisymmetric tensor and subscripts 1 and 2 represent the two spins in the dimer. To eliminate unphysical states, a hard-core constraint is imposed on each dimer, ensuring $s^\dagger s + t_{\alpha}^\dagger t_{\alpha} = 1$. Using the equations (\ref{eq:bo_representation_1}), (\ref{eq:bo_representation_2}) along with the constraint and commutation relations, it can be verified that the spin-spin interaction between two spins are

\begin{equation} \label{eq:sisj_intra} S_{1\alpha\vec{r}}\, S_{2\alpha\vec{r}}= -\frac{3}{4}s^\dagger_\vec{r} s_\vec{r}+ \frac{1}{4}t_{\alpha\vec{r}}^\dagger t_{\alpha\vec{r}} \end{equation}
when the two spins belong to the same dimer ($\vec{r} = \vec{r}^{\,\prime}$), the interaction yields the eigenvalues corresponding to singlet and triplet states, as expected for two spin-$\frac{1}{2}$ operators. In contrast, when the spins belong to different dimers ($\vec{r} \neq \vec{r}^{\,\prime}$), the spin--spin interaction can be expressed as,
  \begin{align}
    S_{m \alpha\vec{r}}\,S_{n\alpha\vec{r}^\prime}=\frac{(-1)^{m+n}}{4}\left[t^\dag_{\alpha\vec{r}} t_{\alpha\vec{r}^\prime }s^\dag_{\vec{r}} s_{\vec{r}^\prime}+t^\dag_{\vec{r}\alpha}t^\dag_{\vec{r}^\prime\alpha}s_ls_k+h.c.\right]\nonumber\\-\frac{(-1)^{m+1}}{4}\left[i\epsilon_{\alpha\beta\gamma} t^\dag_{\vec{r}\alpha}t^\dag_{\vec{r}^\prime\beta}t_{\vec{r}^\prime\gamma}s_k+h.c.\right]\nonumber\\-\frac{(-1)^{n+1}}{4}\left[i\epsilon_{\alpha\beta\gamma}t^\dag_{\vec{r}^\prime\alpha}t^\dag_{\vec{r}\beta}t_{\vec{r}\gamma}s_l+h.c.\right]\nonumber\\-\frac{1}{4}\left[t^\dag_{\vec{r}\alpha}t^\dag_{\vec{r}^\prime\alpha}t_{\vec{r}^\prime\beta}t_{\vec{r}\beta}-t^\dag_{\vec{r}\alpha}t^\dag_{\vec{r}^\prime\beta}t_{\vec{r}^\prime\alpha}t_{\vec{r}\beta}\right]
     \label{eq:sisj_inter}
  \end{align}

where, $m,n=1,2$ label the two sites within a dimer. To simplify the triplon analysis, we approximate the singlet background by a mean field, defined through $\langle s^\dagger \rangle = \langle s \rangle = \bar{s}$, where $\bar{s}$ represents the singlet amplitude per dimer. Under this approximation, the first term of Eq.~\eqref{eq:sisj_inter} describes a condensate of singlets forming the dimerized background. Applying Wick's theorem to the remaining interaction terms and performing a quadratic mean-field decoupling, we find that the middle two terms vanish because of the antisymmetric nature of the Levi-Civita tensor. The fourth term generates effective triplet--triplet interactions, corresponding to interacting triplet pairs. In the main analysis, we neglect these quartic interaction terms and retain only the bilinear contributions in the triplet operators, since previous studies have shown that they produce only small quantitative corrections to the phase diagram~\cite{PhysRevB.41.9323,PhysRevB.110.104402}. For completeness, we explicitly examine the effect of the quartic triplet interactions in Appendix~\ref{appendix:quartic}.

This bond operator representation is applied to the model Hamiltonian (\ref{eq:model_hamiltonian}), where a unit cell consists of two sites (one dimer per unit cell), forms a rectangular Bravais lattice and the translational invariance of the system allow us to incorporate the constraint ($s^\dagger s + t_{\alpha}^\dagger t_{\alpha} = 1$) using the Lagranges multiplier by replacing the local chemical potential with a global chemical potential ($\mu$), then the Hamiltonian becomes,

\begin{align}
H =& \left(-\frac{3}{4}J_d\Bar{s}^2-\mu \Bar{s}^2+\mu\right)N_d+\left(\frac{J_d}{4}-\mu\right)
\sum_{\vec{r}} t_{\vec{r}\alpha}^\dagger t_{\vec{r}\alpha}
\nonumber\\
&+\frac{\Bar{s}^2}{4}
\Bigg[
(-J_1+J_2)
\sum_{\vec{r},\vec{r}+\boldsymbol{\delta}_1}
+\left(-\frac{J_1}{2}+J_3\right)
\sum_{\vec{r},\vec{r}+\boldsymbol{\delta}_2}
\Bigg]
\nonumber\\
&\times
\left(
t^\dagger_{\vec{r}\alpha} t_{\vec{r}'\alpha}
+t^\dagger_{\vec{r}\alpha} t^\dagger_{\vec{r}'\alpha}
+\mathrm{h.c.}
\right)
\label{eq:mf_ham_1}
\end{align}

here, $N_d$ is the number of dimers, and $\mathbf{r}$ denotes the position vector of a dimer on the lattice. The interacting neighboring dimers, coupled via exchange interactions, are located at positions $\mathbf{r}+\boldsymbol{\delta}_1$ and $\mathbf{r}+\boldsymbol{\delta}_2$, where the displacement vectors are defined as
\[
\boldsymbol{\delta}_1 = \pm a \hat{y}, \qquad
\boldsymbol{\delta}_2 = \pm 2a \hat{x}.
\]
Thus, each dimer at position $\mathbf{r}$ is connected to four neighboring dimers along the $\hat{x}$ and $\hat{y}$ directions, as illustrated in Fig.~\ref{fig:dimer_label}. Here, $a$ is the lattice constant, which is set to unity in the following calculations.

\begin{figure}[h]
    \centering

%first column
  \begin{tikzpicture}
  \draw[very thick] (0,1.5) -- (1.5,1.5)
;
  \filldraw (0,1.5) circle (2pt) node [below] {j};  % Left solid circle
  \filldraw (1.5,1.5) circle (2pt) node [below] {i};  % Right solid circle

 \vspace{0.5cm}

    \draw[very thick] (0,0) -- (1.5,0)
      node[midway, above] {$\vec{r}-2a\hat{x}$};
    \filldraw (0,0) circle (2pt) node [below] {i};  % Left solid circle
    \filldraw (1.5,0) circle (2pt) node [below] {j};  % Right solid circle

 \vspace{0.5cm} 
 
  \draw[very thick] (0,-1.5) -- (1.5,-1.5)
;
  \filldraw (0,-1.5) circle (2pt) node [below] {j};  % Left solid circle
  \filldraw (1.5,-1.5) circle (2pt) node [below] {i};  % Right solid circle

%second column
    \hspace{0.23cm}

  \draw[very thick] (3,1.5) -- (4.5,1.5)
    node[midway, above] {$\vec{r}+a\hat{y}$};
  \filldraw (3,1.5) circle (2pt)node[below] {j} ;  % Left solid circle
  \filldraw (4.5,1.5) circle (2pt)node[below] {i};  % Right solid circle

 \vspace{0.5cm}

    % Second line with "$l^{th}$" on top
    \draw[very thick] (3,0) -- (4.5,0)
      node[midway, above] {$\vec{r}$};
    \filldraw (3,0) circle (2pt)node[below] {i} ;  % Left solid circle
    \filldraw (4.5,0) circle (2pt)node[below] {j};  % Right solid circle

 \vspace{0.5cm}

      \draw[very thick] (3,-1.5) -- (4.5,-1.5)
    node[midway, above] {$\vec{r}-a\hat{y}$};
  \filldraw (3,-1.5) circle (2pt)node[below] {j} ;  % Left solid circle
  \filldraw (4.5,-1.5) circle (2pt)node[below] {i};  % Right solid circle

%third column
    \hspace{0.23cm}

  \draw[very thick] (6,1.5) -- (7.5,1.5)
    ;
  \filldraw (6,1.5) circle (2pt)node[below] {j} ;  % Left solid circle
  \filldraw (7.5,1.5) circle (2pt)node[below] {i};  % Right solid circle

 \vspace{0.5cm}
 
    % Second line with "$l^{th}$" on top
    \draw[very thick] (6,0) -- (7.5,0)
      node[midway, above] {$\vec{r}+2a\hat{x}$};
    \filldraw (6,0) circle (2pt)node[below] {i} ;  % Left solid circle
    \filldraw (7.5,0) circle (2pt)node[below] {j};  % Right solid circle

 \vspace{0.5cm}

      \draw[very thick] (6,-1.5) -- (7.5,-1.5)
 ;
  \filldraw (6,-1.5) circle (2pt)node[below] {j} ;  % Left solid circle
  \filldraw (7.5,-1.5) circle (2pt)node[below] {i};  % Right solid circle

  \end{tikzpicture}
      \caption{Schematic representation of the four neighboring dimers of a reference dimer located at position $\mathbf{r}$. The neighboring dimers are located at $\mathbf{r} \pm a\hat{y}$ and $\mathbf{r} \pm 2a\hat{x}$.}
    
    \label{fig:dimer_label}

\end{figure}

Using the Fourier transformation and Fourier identities,

    \begin{subequations}
    \begin{align}
        t_{\vec{r}\alpha}=\frac{1}{\sqrt{N_d}}\sum_{\vec{k}}e^{i\vec{k}\cdot\vec{r_i}}t_{\vec{k}\alpha} \\
        t^\dag_{\vec{r}\alpha}=\frac{1}{\sqrt{N_d}}\sum_{\vec{k}}e^{-i\vec{k}\cdot\vec{r_i}}t^\dag_{\vec{k}\alpha}
    \end{align}
    \end{subequations}
    \begin{equation}
        \delta_{\vec{k},\vec{k'}}=\frac{1}{N_d}\sum_{\vec{k}}e^{-i(\vec{k}-\vec{k'})\cdot\vec{r_i}}
    \end{equation}

 where, $\vec{k}$ vectors takes the values from first Brillouin zone, the mean-field quadratic Hamiltonian in the $\vec{k}$-space can be written as,

\begin{align}
H =& \left(-\frac{3}{4}J_d\bar{s}^2-\mu \bar{s}^2+\mu\right)N_d+\left(\frac{J_d}{4}-\mu\right)
\sum_{\vec{k}} t_{\vec{k}\alpha}^\dagger t_{\vec{k}\alpha}
\nonumber\\
&+\frac{\bar{s}^2}{4}
\Big[
(-J_1+J_2)(2\cos k_y)
+\left(-\frac{J_1}{2}+J_3\right)(2\cos 2k_x)
\Big]
\nonumber\\
&\times
\left(
t^\dagger_{\vec{k}\alpha} t_{\vec{k}\alpha}
+t^\dagger_{\vec{k}\alpha} t^\dagger_{-\vec{k}\alpha}
+\mathrm{h.c.}
\right).
\label{eq:mf_ham_2}
\end{align}

After simplification, the Hamiltonian can be written in a compact form,

\begin{equation}
    H=E_0+\sum_{\vec{k}}\left[A_{\vec{k}}t_{\vec{k}\alpha}^\dagger t_{\vec{k}\alpha}+B_{\vec{k}}\left(t_{\vec{k}\alpha}^\dagger t_{\vec{-k}\alpha}^\dagger+t_{\vec{k}\alpha} t_{\vec{-k}\alpha}\right)\right]
    \label{eq:hamiltonian_quad}
\end{equation}
where,
\begin{equation}
    E_0=\left(-\frac{3}{4}J_d\Bar{s}^2-\mu \Bar{s}^2+\mu\right)N_d
\end{equation}
\begin{equation}
    A_{\vec{k}}=\left(\frac{J_d}{4}-\mu\right)+2B_{\vec{k}}
\end{equation}
\begin{equation}
    B_{\vec{k}}=\frac{\Bar{s}^2}{2} \left[(-J_1+J_2)\cos k_y +\left(\frac{-J_1}{2}+J_3\right)\cos 2k_x\right].
\end{equation}

The Hamiltonian~\eqref{eq:hamiltonian_quad} is bought to diagonal form, using Bogoliubov transformation, which mixes the creation and annihilation operators but keeps their commutation intact. We define the following unitary transformation,
\begin{subequations}
    \begin{align}
    \vartheta_{\vec{k}\alpha}=U_{\vec{k}}t_{\vec{k}\alpha}+V_{\vec{k}}t_{\vec{-k}\alpha}^\dagger \\
     \vartheta_{\vec{k}\alpha}^\dagger=U_{\vec{k}} t^\dagger_{\vec{k}\alpha}
+ V_{\vec{k}} t_{-\vec{k}\alpha}
    \end{align}
\end{subequations}
The operators $\vartheta_{\vec{k}\alpha}$ are the bosons, popularly known as \textit{triplons}, and follow the bosonic commutation relation.
The transformation gives the result,
\begin{align}
    \sum_{\vec{k}}\left[A_{\vec{k}}t_{\vec{k}\alpha}^\dagger t_{\vec{k}\alpha}+B_{\vec{k}}\left(t_{\vec{k}\alpha}^\dagger t_{\vec{-k}\alpha}^\dagger+t_{\vec{k}\alpha} t_{\vec{-k}\alpha}\right)\right]\nonumber\\=\sum_{\vec{k}}\left[\omega_{\vec{k}}\vartheta_{\vec{k}\alpha}^\dagger\vartheta_{\vec{k}\alpha}-\frac{3}{2}(A_{\Vec{k}} - \omega_\vec{k})\right].
\end{align}
Now the Hamiltonian~\eqref{eq:hamiltonian_quad} in the terms of quasi bosonic particles can be written as,
\begin{equation}  H=E_G+\sum_{\vec{k}}\omega_{\vec{k}}\vartheta_{\vec{k}\alpha}^\dagger\vartheta_{\vec{k}\alpha}
\end{equation}
where,
\begin{equation}
    E_G=E_0-\frac{3}{2}\sum_{k}(A_{\vec{k}}-\omega_{\vec{k}})
\end{equation}

\begin{equation}
    \omega_{\vec{k}}=\sqrt{A_{\vec{k}}^2-4B_{\vec{k}}^2}
\end{equation}
$\omega_{\vec{k}}$ is the triplon quasi-particle dispersion. These triplons are the elementary excitations of the system. The spectrum $\omega_{\vec{k}}$ provides insights into the behavior of the system, such as the spin gap and the stability of the quantum ground state. The presence or absence of a gap indicates whether the system is in a gapped quantum disordered phase (with no long-range magnetic order) or in a gapless ordered phase (with magnetic order).
The ground energy per site can be written as,
\begin{equation}
    e_g=\frac{E_G}{2N_d}=\frac{1}{2N_d}\left[E_0-\frac{3}{2}\sum_{\vec{k}}(A_{\vec{k}}-\omega_{\vec{k}})\right].
\end{equation}
The self-consistent equations are obtained by minimizing $e_g$ with respect to $\mu$ and $\Bar{s}^2$. The self-consistent equations are,
\begin{equation}
    \Bar{s}^2=\frac{5}{2}-\frac{3}{2N_d}\sum_{\vec{k}}\frac{A_{\vec{k}}}{\omega_{\vec{k}}}
\end{equation}
\begin{equation}
    \mu=-\frac{3}{4}J_d-\frac{3}{4N_d}\left(\frac{J_d}{4}-\mu\right)\sum_{\vec{k}}\frac{\xi_{\vec{k}}}{\omega_{\vec{k}}}
\end{equation}

where, 

\begin{equation}
    \xi_{\vec{k}}=(-J_1+J_2)(\cos {k_y}) +\left(\frac{-J_1}{2}+J_3\right)(\cos 2{k_x}).
\end{equation}

Since a dimerized phase is the direct product of the singlets, the anomalous expectation value of a singlet boson is non-zero, whereas the expectation value of a single triplet boson is zero, and the expectation value of triplet bosons in pair is non-zero, represents that the singlet bosons and triplet bosons in pair condense whereas a single triplet boson does not condense at dimerized phase. Again at magnetic long-range order, the single triplet boson condenses, giving a non-zero expectation value. The kind of magnetic ordering is determined by the wave vector at which the triplet boson condenses. Qualitatively, this problem can be understood as there is a background of singlets with mean singlet amplitude per bond, and a triplet excitation is formed by breaking a singlet bond which can be dispersed through the background of singlets, assisted by the exchange interactions.

For certain values of coupling strengths, the triplon dispersion becomes gapless at a specific wave vector $\vec{Q}$, this causes a singularity in the  self-consistent equation, the system responds to this by condensing triplons at $\vec{Q}$, these ordering wave vectors are $\vec{Q}=(\frac{\pi}{2},\pi)$ and $(0,0)$ for $J_1=-0.83$ and $J_1=2.83$ respectively. The phenomena of occupying a single quantum state by a macroscopic number of triplons leads to the emergence of a nonzero local magnetic moment, signifying that the the system develops long range order.

From the gapless condition, renormalized chemical potential  can be derived, 

\begin{equation}
    \mu=\frac{J_d}{4}+4B_\vec{Q}.
\end{equation}
In the ordered phase, the triplon density $n_c$ can be defined as the average number of condensed triplons per dimer

    \begin{equation}
        n_c=\frac{1}{N_d}\langle t_{\vec{Q}\alpha}^\dag t_{\vec{Q}\alpha}\rangle.
    \end{equation}

To determine the self-consistent parameters of the system, the total triplon density is split into two parts: one for $\vec{k}=\vec{Q}$ (where triplon condensation occurs) and one for $\vec{k}\neq \vec{Q}$. Since there are two wave vectors where condensation occurs ($\vec{Q}=(\frac{\pi}{2},\pi)$ and $(0,0)$), the condensation density is sum over these two modes, and the triplon condensation density is given by,

    \begin{equation}
        n_c=\frac{1}{N_d}\langle t_{\vec{Q}\alpha}^\dag t_{\vec{Q}\alpha}\rangle=1-s^2-\frac{1}{N_d}\sum_{\vec{k}\neq \vec{Q}}\langle t_{\vec{k}\alpha}^\dag t_{\vec{k}\alpha}\rangle.
    \end{equation}
 After doing a Bogoliubov transformation as done before, the self-consistent equations for the ordered phases given by,
\begin{equation}
    {\bar{s}}^2=\frac{5}{2}-n_c-\frac{3}{2N_d}\sum_{\vec{k}\neq \vec{Q}}\frac{A_\vec{k}}{\omega_\vec{k}}
\end{equation}

\begin{equation}
    n_c=\frac{1}{\xi_\vec{Q}}\left[\mu+\frac{3}{4}J_d-\frac{3}{4N_d}\left(\frac{J_d}{4}-\mu\right)\sum_{\vec{k}\neq \vec{Q}}\frac{\xi_\vec{k}}{\omega_\vec{k}}\right]
\end{equation}

where, ${\vec{Q}}=({Q_x},{Q_y})$ can take values $(\frac{\pi}{2},\pi)$ and $(0,0)$, and,
 \begin{equation}
     \xi_{\vec{Q}}=(-J_1+J_2)(\cos {Q_y}) +\left(\frac{-J_1}{2}+J_3\right)(\cos 2{Q_x}).
 \end{equation}

This field theoretical method provides a convenient framework for describing dimerized quantum spin systems; however, it has several inherent limitations. In particular, it is known to overestimate the spin gap and cannot reliably establish the coexistence of magnetic and dimer orders, as discussed in previous studies~\cite{brenig1997,bouzerar2001,hwang2012,doretto2014}. Furthermore, the approach is biased by the choice of a dimerized reference state and may therefore overemphasize singlet-dominated phases or fail to capture competing orders. Despite these limitations, the method provides valuable qualitative insights into the stability of dimerized phases and the overall structure of the phase diagram. To overcome these shortcomings, we complement our analysis with DMRG calculations, which allow for a more accurate and unbiased investigation of the phases in the system.

% section bond_operator_mean_field_theory (end)
\section{Results and discussion} % (fold)
\label{sec:results_and_discussion}

\begin{figure}
  \centering
    \includegraphics[width=0.98\columnwidth]{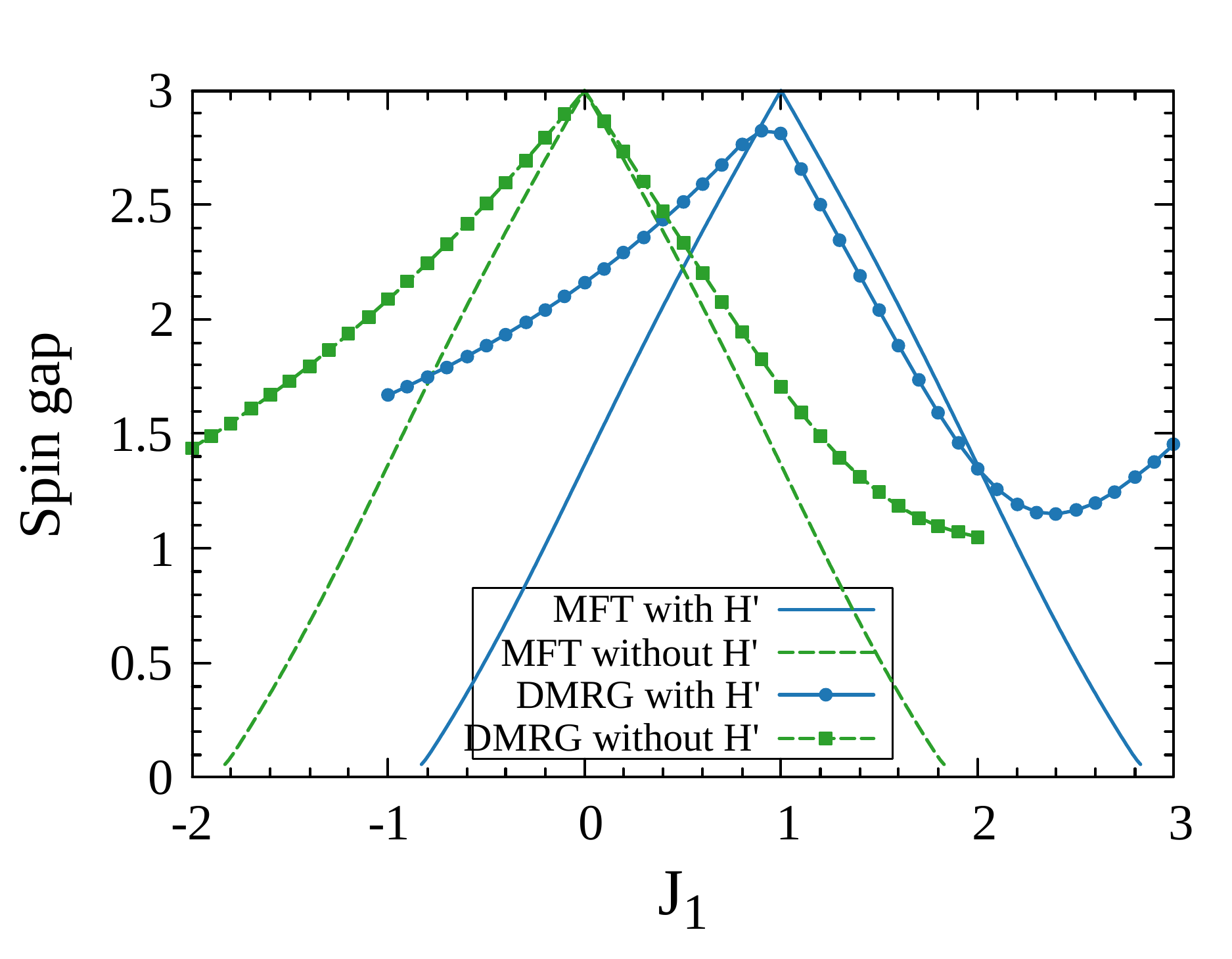}

  \caption{Spin gap obtained from mean-field theory and DMRG ($8\times6$) for the frustrated and unfrustrated systems. In both cases, the gap is maximum at the exactly solvable points and decreases on either side.}
  \label{fig:structure_factor}
\end{figure}

\begin{figure}
    \centering
    \includegraphics[width=0.98\columnwidth]{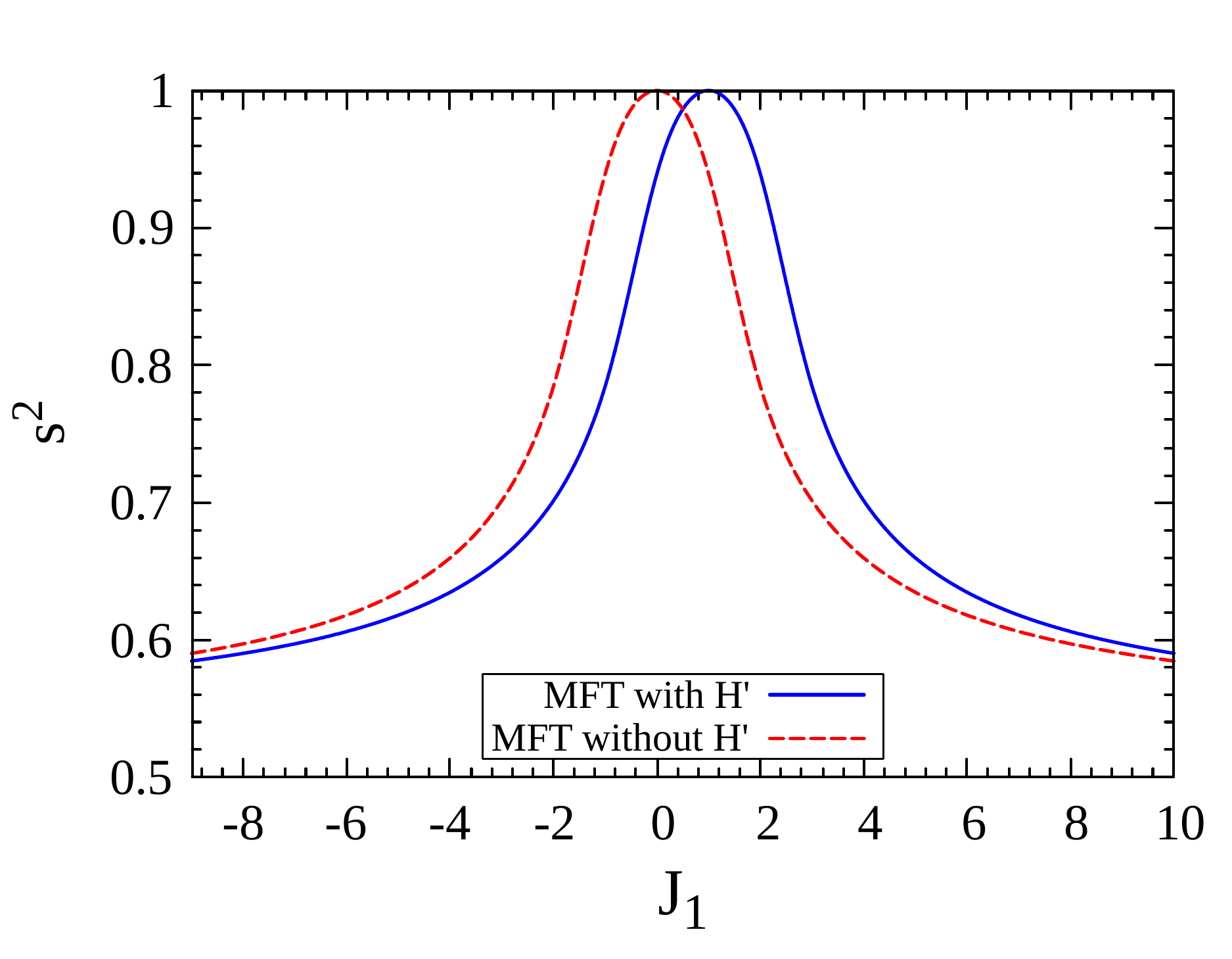}
    \caption{This figure illustrates the singlet condensation density on dimers for the case of frustrated and unfrustrated lattice.}
    \label{fig:s2}
\end{figure}

\begin{figure}
  \centering
    \includegraphics[width=0.95\columnwidth]{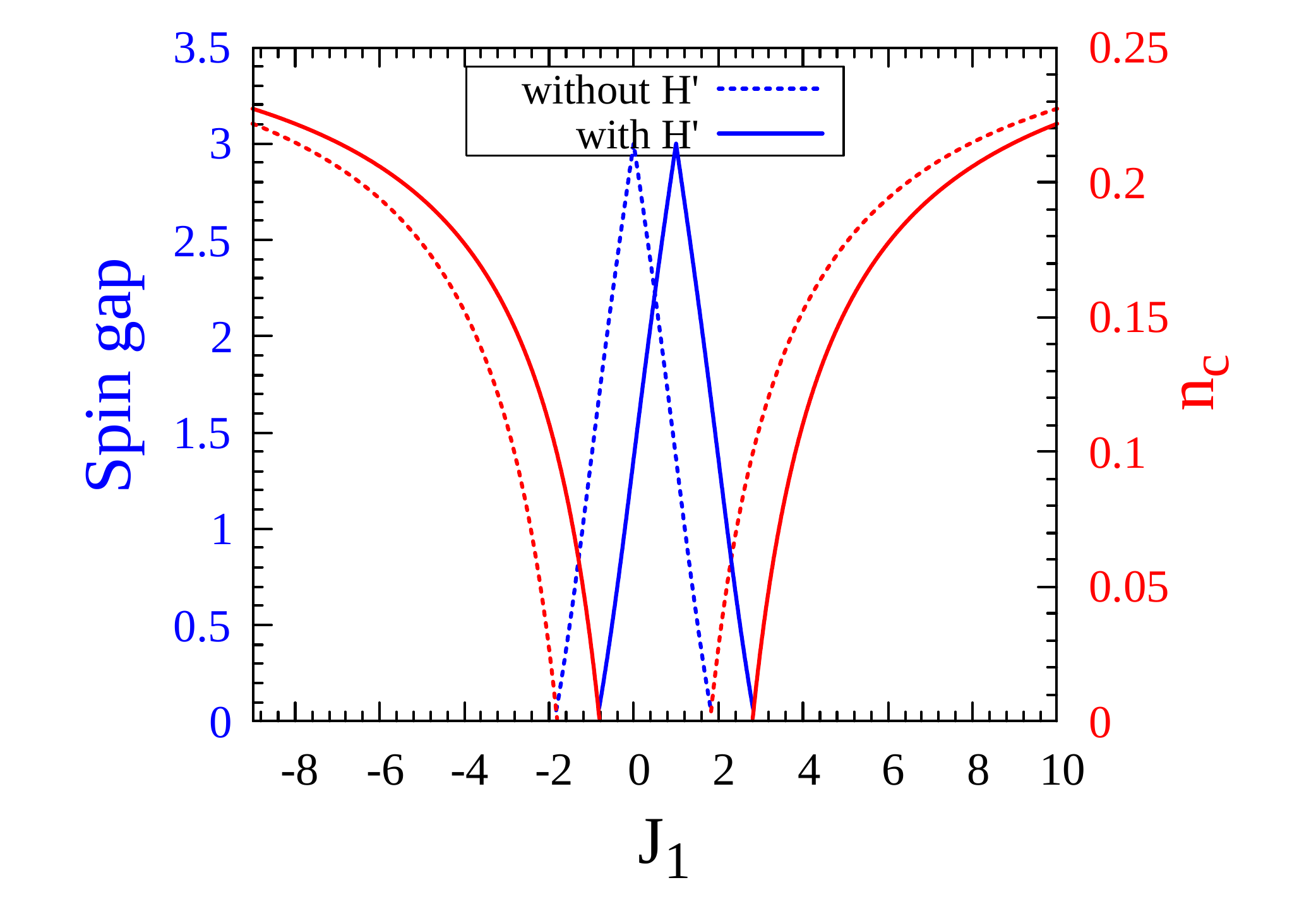}
  \caption{This figure illustrates the spin gap ($\omega_\vec{k}$) obtained from the mean-field theory and triplet condensation density with different scales on the y-axis. Dashed lines are for the system without frustration, and solid line is for the model with frustrated interactions. The ordering wave vector for which the gap closes and $n_c$ increases is same for both cases, i.e. ($\frac{\pi}{2},\pi$) and ($0,0$). }
  \label{fig:gap_nc}
\end{figure}

\begin{figure}
  \centering
    \includegraphics[width=.23\textwidth]{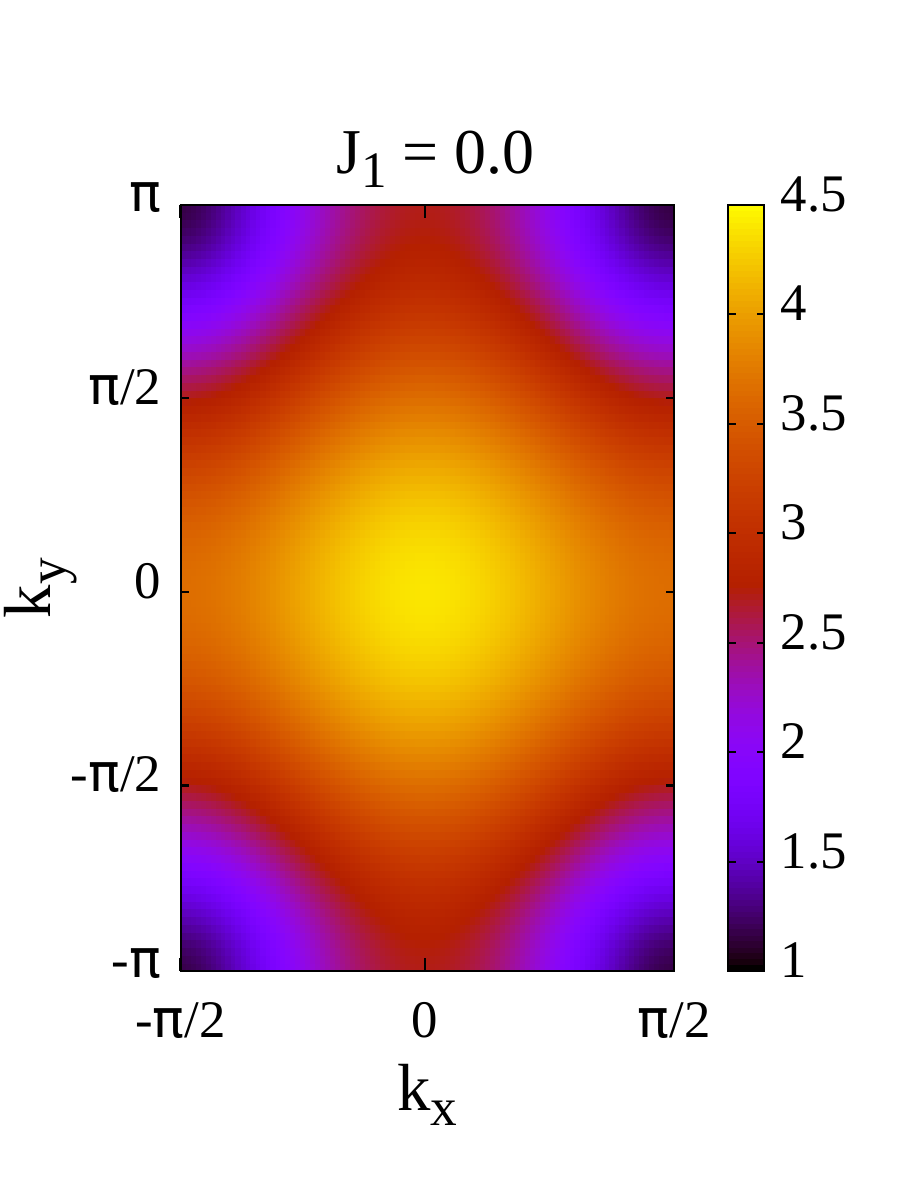}
    \includegraphics[width=.23\textwidth]{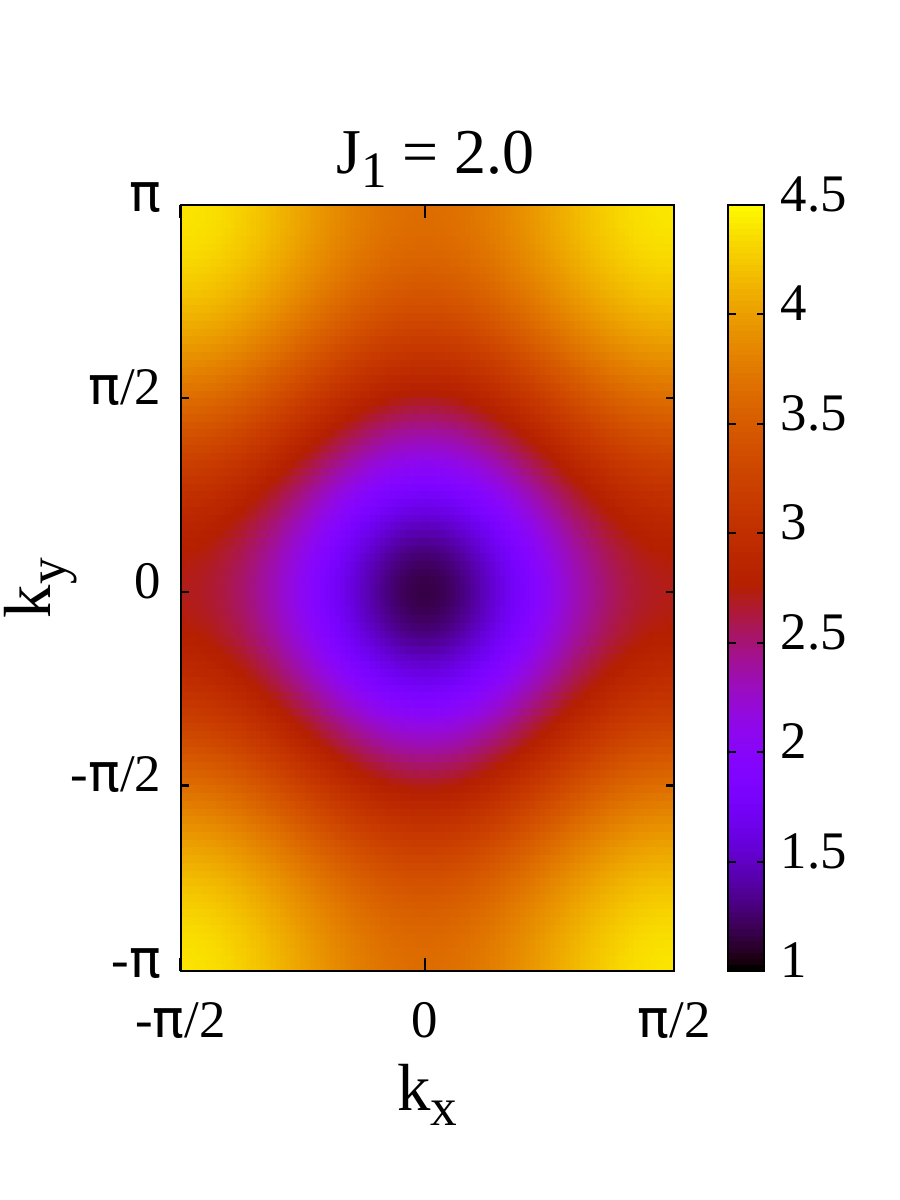}
  \caption{Dispersion of quasi-particles in the first Brillouin zone, obtained from BOMFT, for the double-period stripe phase at $J_1=0.0$ and the Néel-ordered phase at $J_1=2.0$.}
  \label{fig:disp}
\end{figure}

\begin{figure}
    \centering
    \includegraphics[width=0.99\columnwidth]{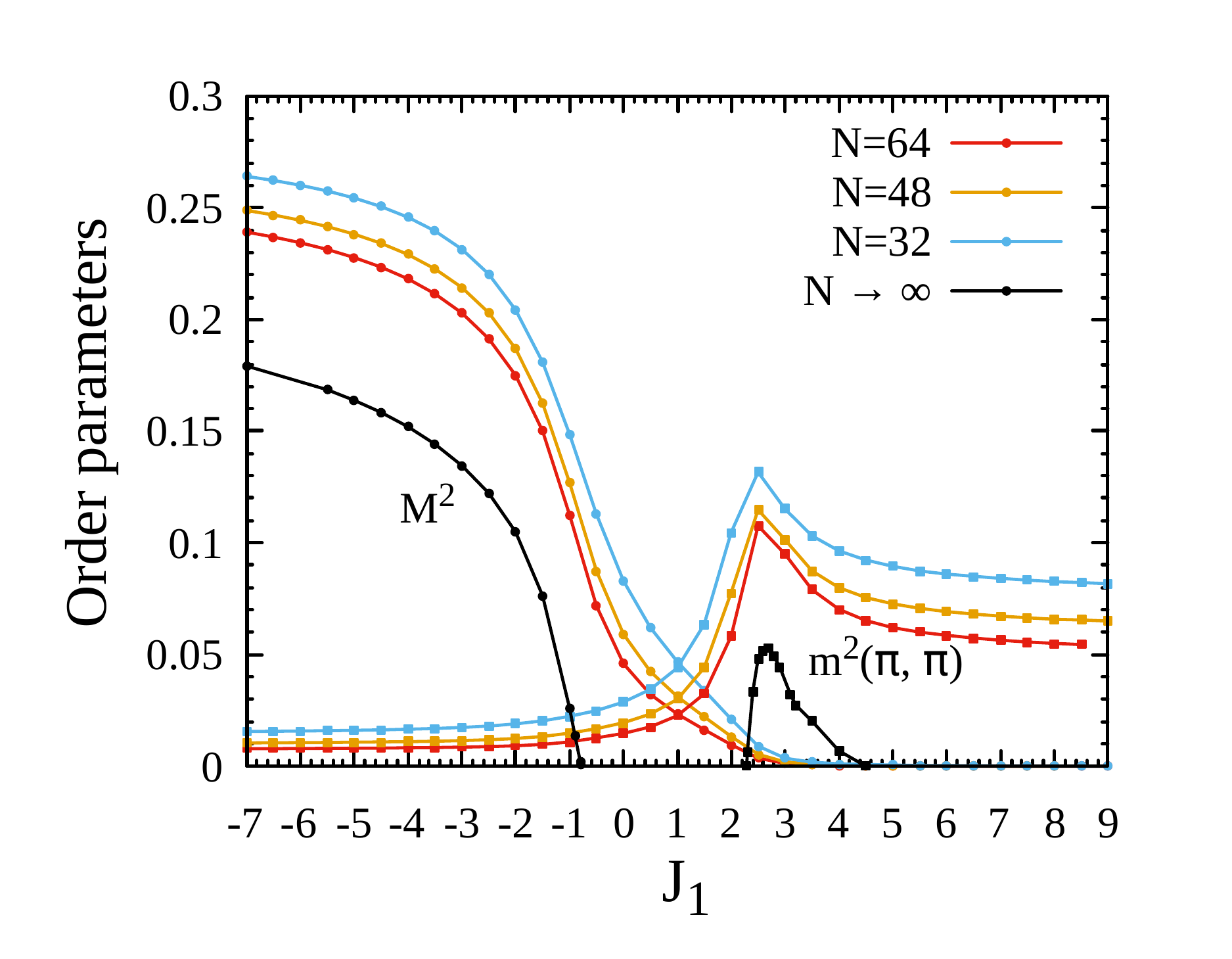}
    \caption{Order parameters for the Néel phase ($m^2(\pi,\pi)$) and the double-stripe phase ($M^2$). Curves with filled circular markers correspond to the double-stripe order, while those with square markers represent the Néel order. Results for different system sizes are indicated by color: red for $8\times8$, brown for $8\times6$, and blue for $8\times4$. The black curves denote the extrapolated values in the thermodynamic limit.}
    \label{fig:sub_mag}
\end{figure}

\begin{figure*}[t]
 \centering
    \includegraphics[width=.245\textwidth]{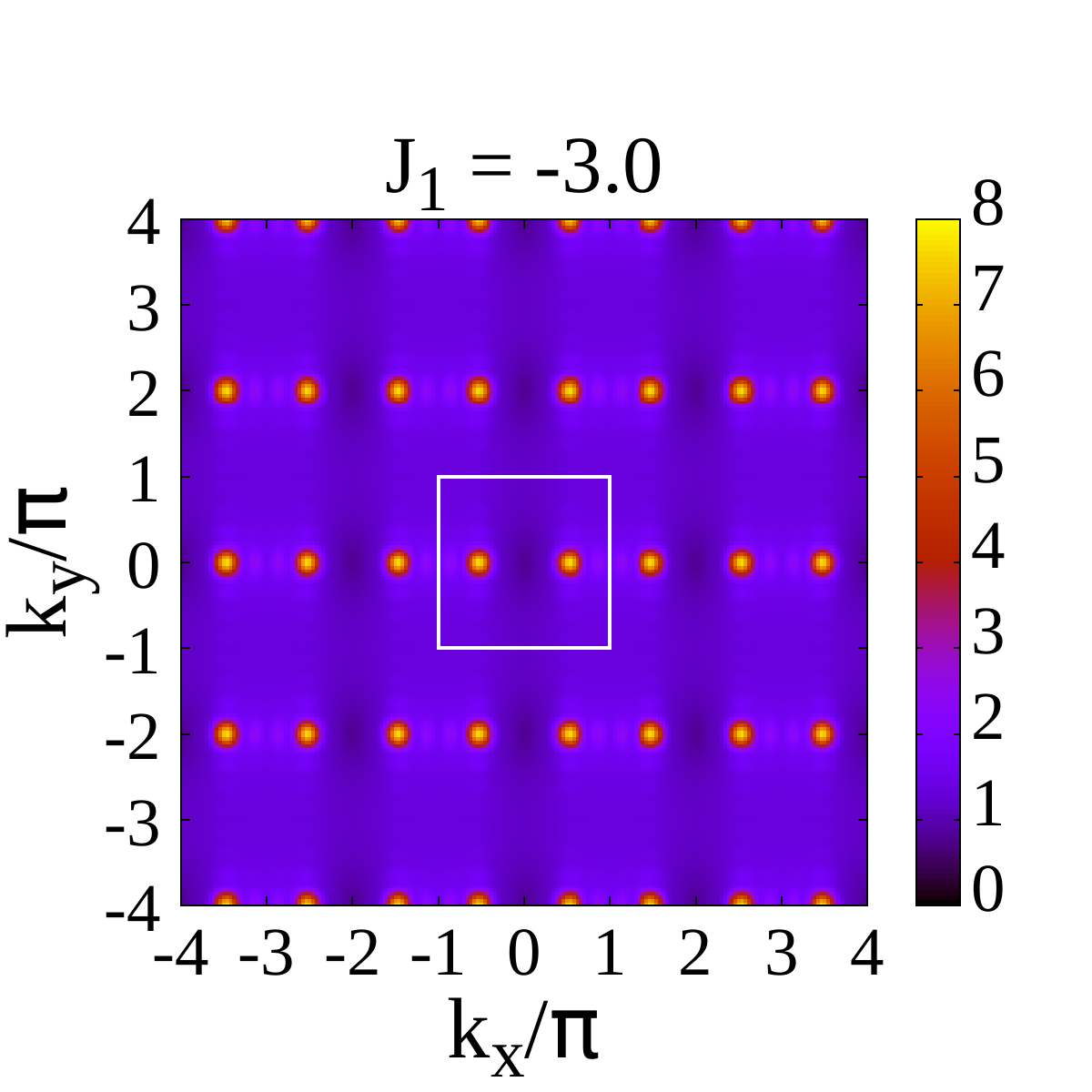}
   \includegraphics[width=.245\textwidth]{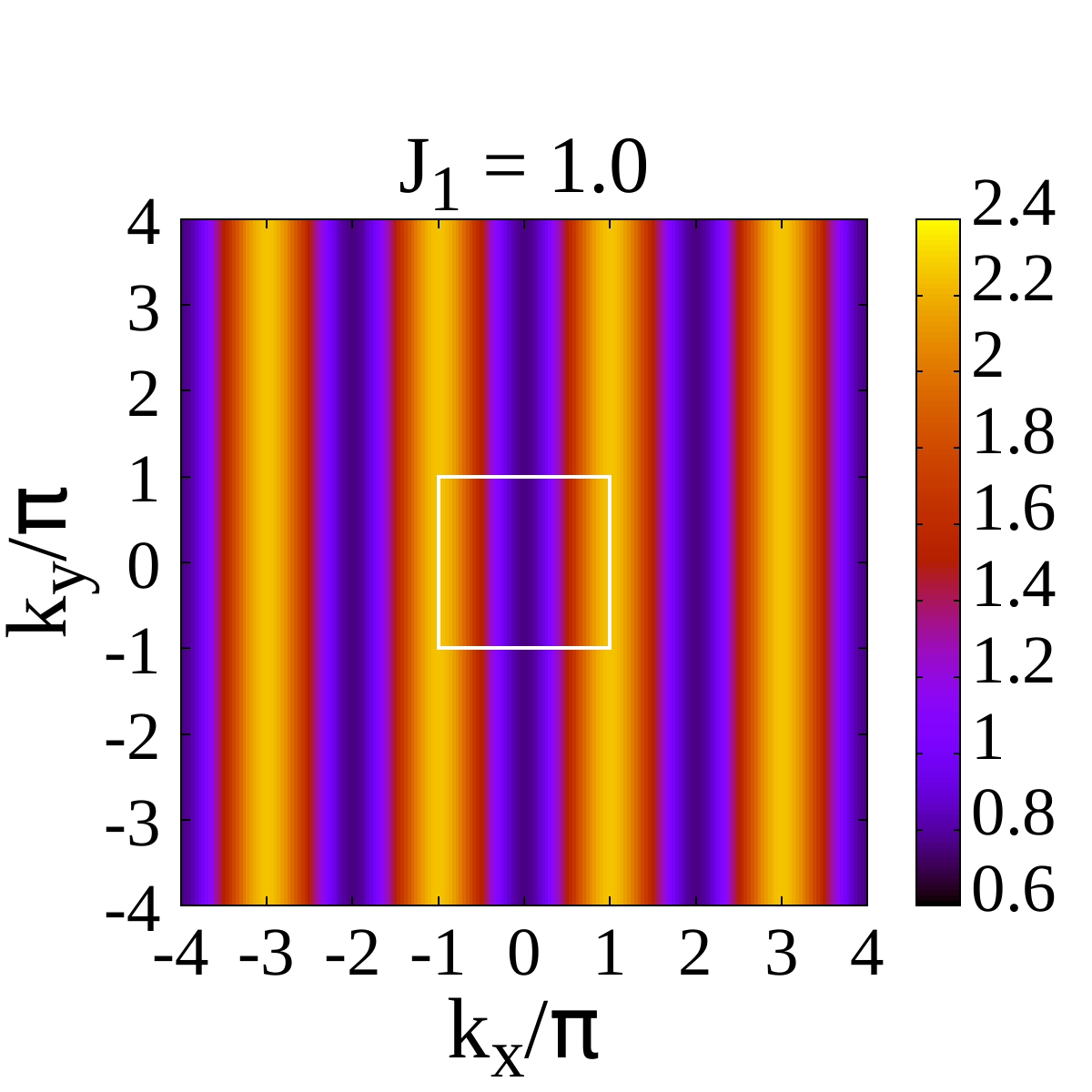}
   \includegraphics[width=.245\textwidth]{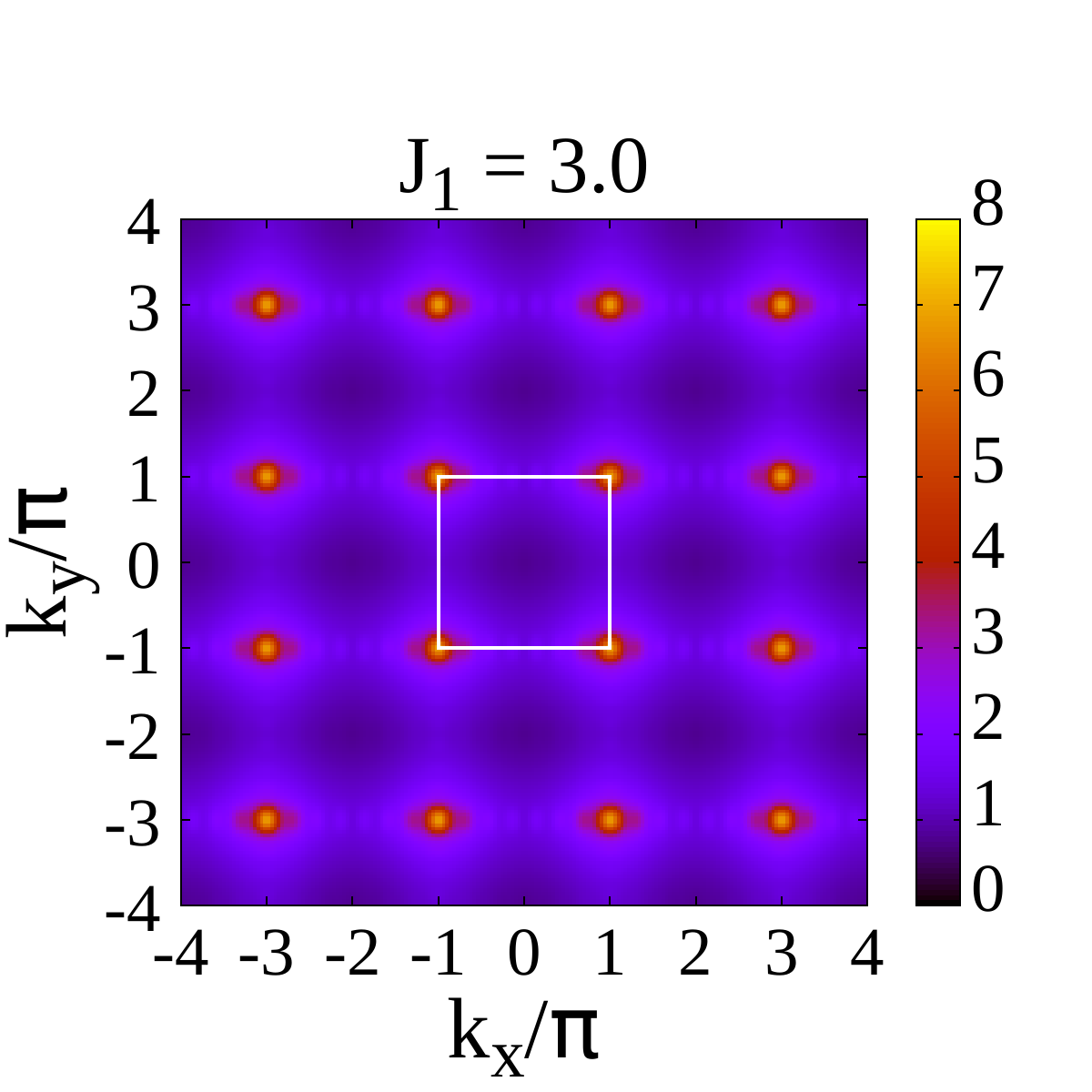}
   \includegraphics[width=.245\textwidth]{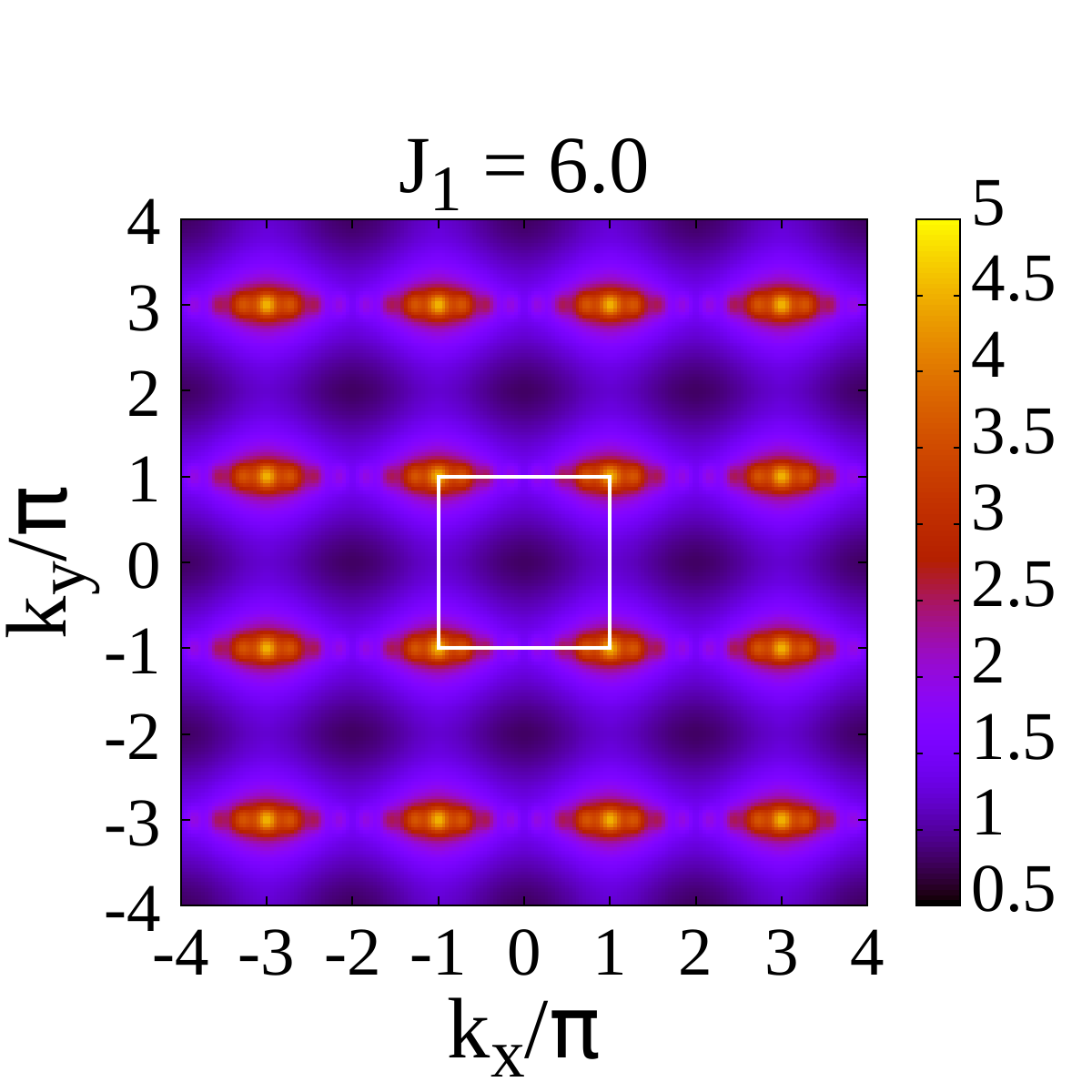}    
  \caption{The static spin structure factor derived from DMRG for double period striped phase at $J_1=-3.0$ and Neel order at $J_1=3.0$, product singlet state at $J_1=1$ and the broad peak structure observed at $J_1=6.0$. The first Brillouin zone is marked by the white-lined square.}
  \label{fig:str_fact}
\end{figure*}

\begin{figure}
    \centering
    \includegraphics[width=0.4\textwidth]{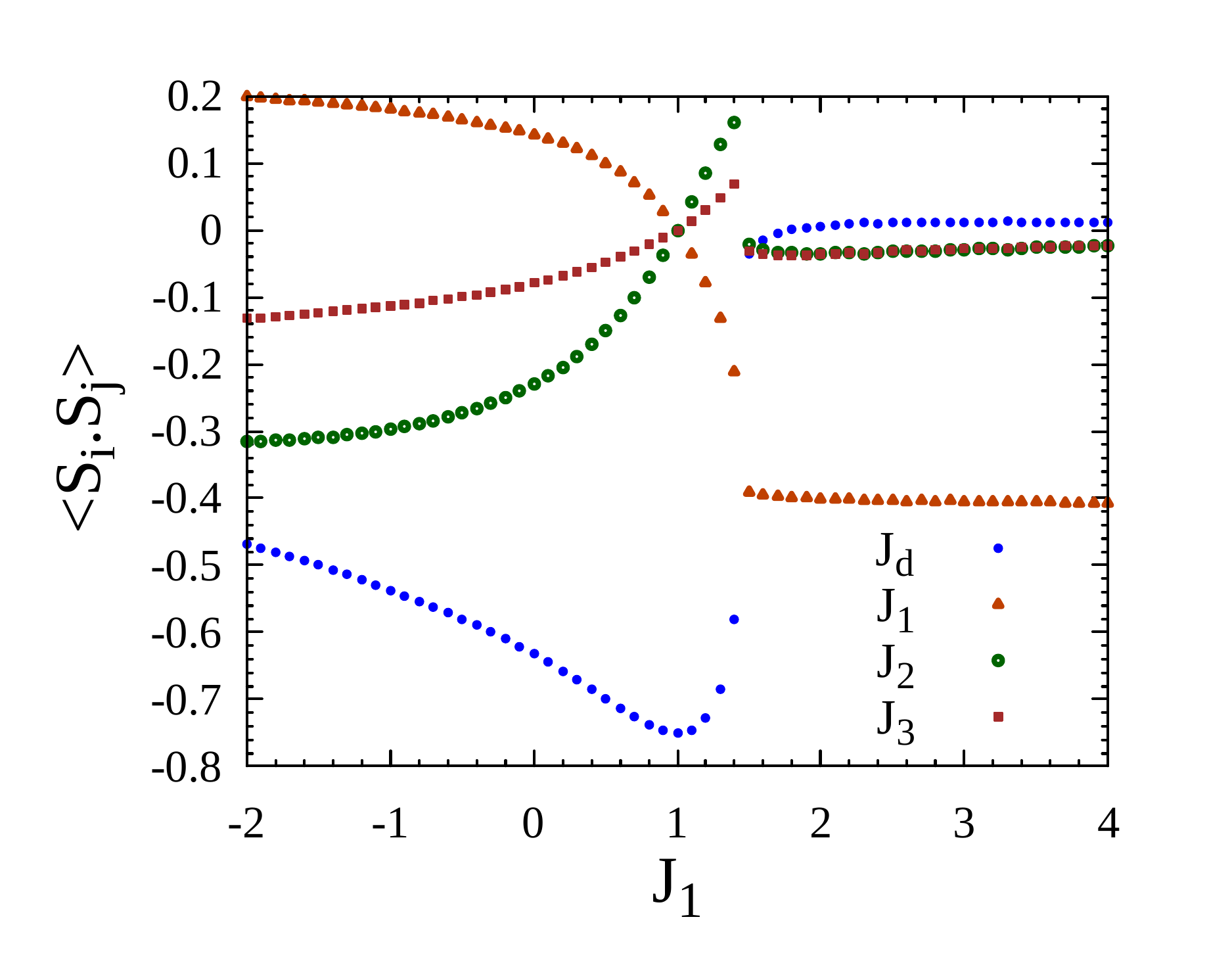}
    \caption{Average spin--spin correlations for different bond types as a function of $J_1$, obtained by averaging the correlations over all bonds of a given type.}
    \label{fig:av_bond}
\end{figure}

In this section, we present and analyze the results obtained from the BOMFT and DMRG calculations. The DMRG simulations were performed using the ITensors library~\cite{itensor}. For DMRG cluster, we have considered a cylindrical boundary condition, along the y-direction the system is open and along x-direction the boundary is closed and periodic, this periodicity stabilizes the product singlet states. The calculations were carried out on system sizes $8\times4$, $8\times6$, and $8\times8$. A maximum of $m=600$ states were retained, with the truncation error kept below $10^{-5}$. Within the BOMFT framework, various physical quantities were computed by numerically solving the self-consistent equations corresponding to both the ordered and disordered phases.

 Using the bond-operator mean-field approach, we calculate the spin gap and investigate its behavior as a function of the nearest-neighbor interaction strength, $J_1$ (Fig.~\ref{fig:structure_factor}). Without frustration (without $H'$ term), the model reduces to a nearest-neighbor Heisenberg Hamiltonian. As a result, at $J_1 = 0$, the system becomes a set of decoupled dimers and the elementary excitation corresponds to breaking a singlet into a triplet on an isolated bond. In the frustrated case (with $H'$ term), the dimerized state is stabilized as the ground state due to an intricate interplay between quantum fluctuations and frustration. At the exact point ($J_1=1$), the excitation spectrum is dominated by localized triplon excitations created by breaking a singlet dimer. The energy required to convert a singlet bond into a triplet is exactly $J_d$, since the singlet bond minimizes the exchange energy on the corresponding block triangles. Once the singlet is broken, this local energy gain is lost, leading to a triplon excitation cost of precisely $J_d$. Thus, the spin gap reaches $\omega_{\vec{k}}=J_d=3$ at $J_1=1$. As $J_1$ is tuned away from this point, interdimer interactions become increasingly important, allowing the triplons to delocalize and lowering the excitation energy. Consequently, the spin gap decreases from its maximum value. Fig.~\ref{fig:structure_factor} shows that the spin-gap values obtained from DMRG are in good agreement with the bond-operator mean-field results. The agreement is particularly accurate in the decoupled-dimer limit (dotted curve), where the triplon picture is essentially exact. Near the exactly dimerized point (solid curve), a small deviation between the two approaches is observed, which can be attributed primarily to finite-size effects in the DMRG calculations.

The quantity $\bar{s}^2$, derived from the self-consistent equations, measures the expectation value of the singlet projection operator $\left(\frac{1}{4} - \vec{S}_1 \cdot \vec{S}_2\right)$ on a dimer in the mean-field dimerized ground state. It reaches its maximum value of $\bar{s}^2=1$ at $J_1 = 0$ and $J_1 = 1$ for the frustrated and unfrustrated cases, respectively, as shown in Fig.~\ref{fig:s2}, showing a full condensation of singlets. As $J_1$ is varied, $\bar{s}^2$ decreases on both sides but remains finite throughout the parameter range, indicating that the mean singlet amplitude of the system stays nonzero.

The mean-field ground state is a quantum-disordered phase when the triplons are gapped and exhibit zero magnetic moment. However, an ordered phase begins to emerge at a certain ordering wavevector $\vec{Q}$ when the spin gap closes. Fig.~\ref{fig:gap_nc} shows the triplet condensation density ($n_c$) alongside the spin gap. It can be seen that the triplet condensation density starts to increase from zero as the spin gap vanishes, signaling the emergence of two ordered phases with ordering wavevectors $\left(\frac{\pi}{2},\pi\right)$ and $(0,0)$, respectively. These ordering wave vectors are also evident from the dispersion plot shown in Fig.~\ref{fig:disp}, as one can see, the dispersion is minimum at $\left(\frac{\pi}{2},\pi\right)$ for $J_1=0.0$ and $(0,0)$ for $J_1=2.0$. $n_c$ stays zero in the region $-0.81<J_1<2.81$, showing that there are only singlets on the bonds in the ground state, so this region is a quantum disordered dimerized phase.

The long-range ordered phases in the system can be identified by examining the wave vectors \( \vec{Q} \), for a lattice with two sites per unit cell (one dimer per unit cell), associated with a columnar dimer lattice and sublattice labeling (Fig.~\ref{fig:dimer_label}). Specifically, the wave vectors \( \vec{Q} = \left( \frac{\pi}{2}, \pi \right) \) and \( \vec{Q} = (0, 0) \) correspond to two distinct types of magnetic order:
\begin{enumerate}
    \item \emph{Néel Antiferromagnetic Order}: This phase is characterized by alternating up-and-down spin configurations on a bipartite lattice. It typically arises for wave vectors of the form \( \vec{Q} = (0, 0) \), indicating that the spin correlation between neighboring sites alternates over the lattice.
    \item \emph{Double-period Stripe Order}: The wave vector \( \vec{Q} = \left( \frac{\pi}{2}, \pi \right) \) leads to a double-period stripe ordering, where spins alternate in blocks of two columns. Specifically, the first and second columns exhibit up spins, the third and fourth columns show down spins, and this pattern repeats periodically. This results in a stripe-like structure with a doubled periodicity, where the modulation of spins repeats after every two columns.
\end{enumerate}

To further investigate the emergence of ordered phases and accurately determine the critical points, we introduce several order parameters tailored for finite-size clusters. The Néel order parameter, associated with the antiferromagnetic phase, can be derived using a \( \vec{k} \)-dependent magnetic susceptibility as described in Ref.~\cite{Schulz_1996}. It is given by the expression:
\begin{equation}
    m^2(\pi, \pi) = \frac{1}{N(N+2)} \sum_{i,j} \langle \vec{S}_i \cdot \vec{S}_j \rangle e^{\iota \vec{k} \cdot (\vec{r}_i - \vec{r}_j)},
    \label{neel}
\end{equation}
where \( \vec{k} = (\pi, \pi) \), $N$ is the number of sites and \( \vec{r}_i \) is the position of the \( i \)-th spin.  

For the double-period stripe phase, we define an order parameter by considering unit cells consisting of four sites. The order parameter is expressed as:
\begin{equation}
    M^2 = \frac{1}{16N_{\text{uc}}^2} \sum_{R,R'} \sum_{i,j} (-1)^{i+j} \langle \vec{S}_i(R) \cdot \vec{S}_j(R') \rangle,
    \label{stripe}
\end{equation}
where $N_{uc}$ is the number of unit cells, \( R \) and \( R' \) denote the positions of unit cells, and \( i, j \) are the indices of sites within a unit cell. To simplify the definition, we consider a unit cell consisting of a single site. Within this framework, we assign labels such that even-indexed sites correspond to up spins and odd-indexed sites to down spins. The resulting order parameter for this alternative labeling scheme is expressed as:
\begin{equation}
    M^2 = \frac{1}{N^2} \sum_{i,j} (-1)^{i+j} \langle \vec{S}_i \cdot \vec{S}_j \rangle.
    \label{stripe_alt}
\end{equation}

Fig.~\ref{fig:sub_mag} illustrates these order parameters. It is evident that the double-stripe order parameter vanishes for strong antiferromagnetic values of \( J_1 \), while the Néel order parameter vanishes for strong ferromagnetic \( J_1 \). Interestingly, these two order parameters intersect precisely at the exact point for finite sizes. Notably, the double-stripe order parameter exhibits a significantly higher magnitude ($M^2\approx 0.25$), indicating a strong and robust order in this phase. In contrast, the Néel order parameter has a much smaller magnitude ($m^2(\pi,\pi)\approx 0.07$) for large $J_1$, suggesting the possibility of an additional phase emerging at large \( J_1 \), which warrants further investigation.

To further investigate the nature of the ordered phases in our system, we calculate the static structure factor, which is a key quantity for probing the long-range correlations and spatial ordering of spins. The static structure factor \( S(k) \) is given by the Fourier transform of the spin-spin correlation function:

\begin{equation}
    S(k) = \frac{1}{N} \sum_{i,j=1}^{N} \langle \vec{S}_i \cdot\vec{S}_j \rangle e^{\iota \vec{k} \cdot (\vec{r}_i - \vec{r}_j)},
    \label{eq:str_fct}
\end{equation}
where \( \vec{r}_i \) and \( \vec{r}_j \) are the position vectors of the spins at sites \( i \) and \( j \), respectively. The structure factor is an important tool for identifying the ordering wavevectors and detecting different phases. Peaks in the structure factor correspond to the wavevectors at which spin correlations are enhanced, indicating the presence of long-range order.

Fig.~\ref{fig:str_fact} displays the structure factor for different values of \( J_1 \). At \( J_1 = 3.0 \), a peak at \( (\pi, \pi) \) clearly indicates Néel long-range order, which is characteristic of antiferromagnetic alignment. On the other hand, for \( J_1 = -3.0 \), a peak appears at \( (\frac{\pi}{2}, 0) \), signaling the formation of a double-period stripe order. As we increase \( J_1 \), the peaks begin to broaden and lose intensity, with the magnitude of the peak reducing to approximately 5 at \( J_1 = 6.0 \), as shown in the right plot of Fig.~\ref{fig:str_fact}. This suggests the onset of a quantum-disordered phase, where no well-defined long-range order is present. Moreover, at \( J_1 = 1.0 \), the structure factor shows less intense broaden peaks, further confirming the presence of a quantum-disordered phase at the exact point.

Finite-size effects are clearly visible in the order parameters, as evidenced by their systematic dependence on system size (see Fig.~\ref{fig:sub_mag}). In particular, smaller systems tend to overestimate the magnitude of the order parameter, and the transition region appears broadened and slightly shifted. As the system size increases, the transition becomes sharper and the results converge toward the thermodynamic limit, providing a more reliable estimate of the phase boundaries. 

To further quantify the phase transitions, we performed a finite-size scaling analysis of both the double-stripe and N\'eel order parameters. In particular, we carried out least-squares fits using data from cluster sizes $8\times4$, $8\times6$, and $8\times8$, which allowed us to extract the corresponding critical points in the thermodynamic limit. A detailed description of the finite-size scaling procedure is provided in Appendix~\ref{appendix:finite_size}. The extrapolation to the thermodynamic limit indicates that the double-stripe order parameter vanishes at \( J_1 \approx -0.79 \). A similar finite-size scaling analysis was conducted for the Néel order parameter. Remarkably, the Néel order persists only within a narrow range of \( 2.29 < J_1 < 4.5 \). This observation is consistent with the results depicted in Fig.~\ref{fig:str_fact}, where the structure factor at \( J_1 = 6.0 \) reveals a broadening peak. This broadening is indicative of an additional phase. Importantly, this phase remains robust as \( J_1 \) increases further. This behaviour is understood by calculating the averaged spin-spin correlation on all types of bonds present in the system as shown in Fig.~\ref{fig:av_bond}, which shows that for large positive $J_1$ the average correlation of every bond vanishes except for $J_1$ bonds, indicating a antiferromagnetic correlation.

\section{Conclusions}
\label{sec:conclusions}
In this work we have introduced a frustrated spin-½ Heisenberg model on a coupled ladder and demonstrated that, for a specific ratio of exchange couplings, the Hamiltonian admits an exact columnar dimer ground state. Our combined BOMFT and DMRG analysis reveals a sequence of quantum phases controlled by $J_1$. Within BOMFT, the triplon gap closes at $J_1 = -0.81$ and $J_1 = 2.81$, signaling transitions from the double-period stripe phase to the dimerized phase and from the dimerized phase to the Néel phase, respectively. DMRG refines these boundaries, locating the stripe–dimer transition at $J_1 \approx -0.79$ and the dimer–Néel transition at $J_1 \approx 2.29$. The columnar dimer phase therefore remains stable in the window $-0.79 < J_1 < 2.29$, demonstrating the robustness of the singlet product state. The Néel phase persists only within a finite interval $2.29 < J_1 < 4.5$, beyond which long-range order is suppressed and the system crosses over into a quasi-one-dimensional regime dominated by antiferromagnetic correlations along the $J_1$ bonds. Overall, these results establish a rich phase diagram and highlight the strong quantitative consistency between mean-field theory and DMRG. The identification of an exact ground state within this geometry offers a rare theoretical benchmark that may guide future studies of exotic quantum phases in frustrated spin ladders and related low-dimensional magnetic materials. We also presented an alternative representation of the model as a network of orthogonal zigzag and fully frustrated spin ladders, offering a structural framework conducive to quantum materials engineering.

\section{Acknowledgements}
\label{sec:acknowledgements}
Manas Ranjan Mahapatra acknowledges the financial support from University
Grant Commission (UGC), New Delhi, India.

\appendix

\section{Effect of triplet interactions in BOMFT}
\label{appendix:quartic}

\begin{figure}
    \centering
    \includegraphics[width=0.88\columnwidth]{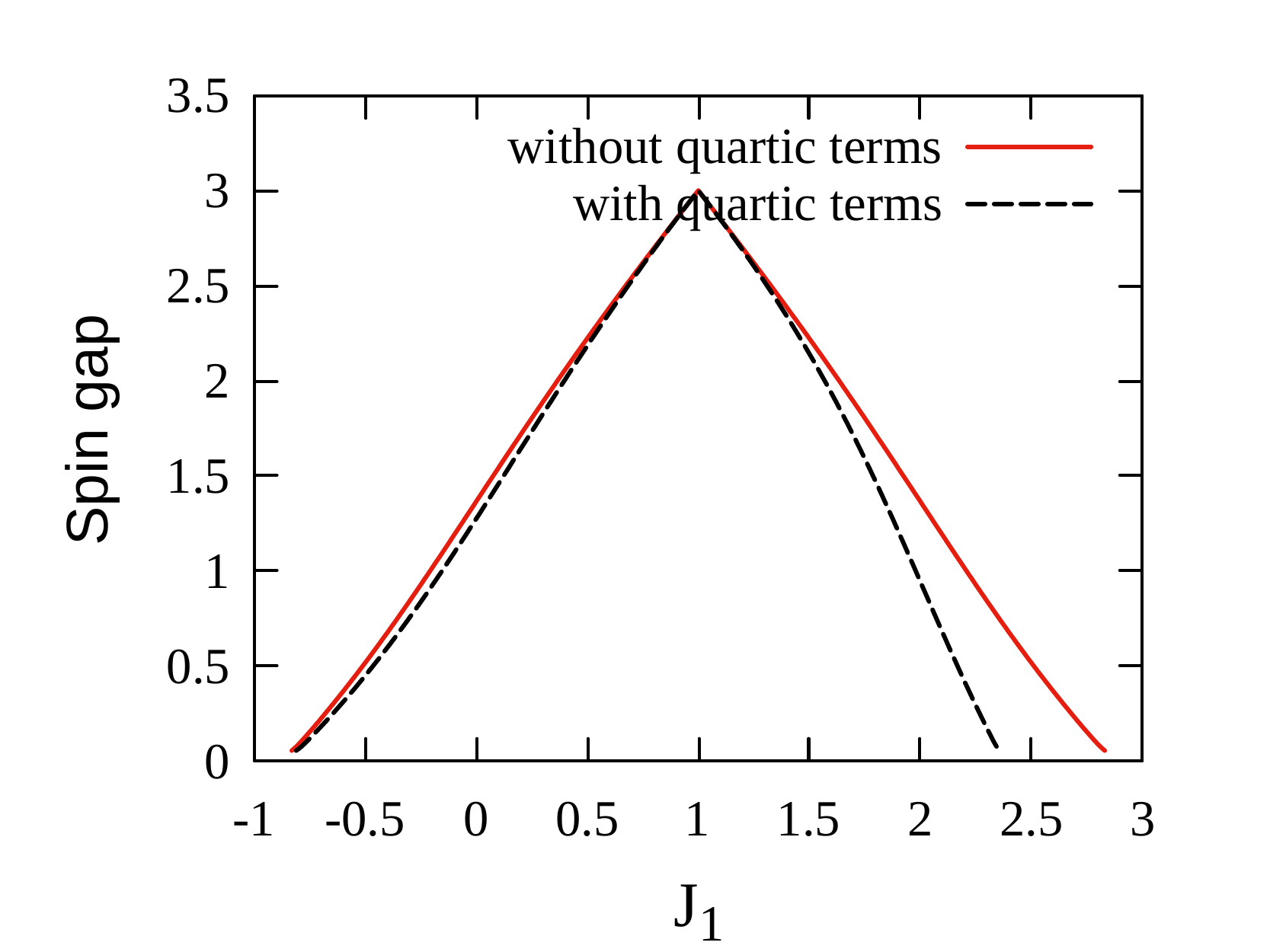}
    \caption{Spin gap obtained within the harmonic approximation and with quartic triplet--triplet interactions included through mean-field decoupling.}
    \label{fig:trip_int}
\end{figure}

In the harmonic approximation discussed in the main text, only the bilinear contribution arising from the first term of Eq.~\eqref{eq:sisj_inter} was retained. To examine the effect of triplet--triplet interactions, we additionally include the quartic terms through a mean-field quadratic decoupling scheme.

Introducing the mean-field parameters
\begin{equation}
    P=\langle t_{\vec{r}\alpha}^\dagger t_{\vec{r}^{\,\prime}\alpha}\rangle,
\end{equation}
and
\begin{equation}
    Q=\langle t_{\vec{r}\alpha}^\dagger t_{\vec{r}^{\,\prime}\alpha}^\dagger\rangle,
\end{equation}
the Hamiltonian including quartic corrections can be written as

\begin{align}
H =&
\left(
-\frac{3}{4}J_d\bar{s}^2
-\mu \bar{s}^2
+\mu
\right)N_d
+
\left(
\frac{J_d}{4}-\mu
\right)
\sum_{\vec r}
t_{\vec r\alpha}^\dagger
t_{\vec r\alpha}
\nonumber\\[4pt]
&+
\frac{\bar{s}^2}{4}
\Bigg[
(-J_1+J_2)
\sum_{\vec r,\vec r+\boldsymbol{\delta}_1}
+
\left(
-\frac{J_1}{2}+J_3
\right)
\sum_{\vec r,\vec r+\boldsymbol{\delta}_2}
\Bigg]
\nonumber\\
&\qquad \times
\left(
t_{\vec r\alpha}^\dagger
t_{\vec r^{\,\prime}\alpha}
+
t_{\vec r\alpha}^\dagger
t_{\vec r^{\,\prime}\alpha}^\dagger
+\hc
\right)
\nonumber\\[4pt]
&-
\frac{1}{4}
\Bigg[
(J_1+J_2)
\sum_{\vec r,\vec r+\boldsymbol{\delta}_1}
+
\left(
\frac{J_1}{2}+J_3
\right)
\sum_{\vec r,\vec r+\boldsymbol{\delta}_2}
\Bigg]
\nonumber\\
&\qquad \times
\Big[
Q
\left(
t_{\vec r\alpha}^\dagger
t_{\vec r^{\,\prime}\alpha}^\dagger
+
t_{\vec r\alpha}
t_{\vec r^{\,\prime}\alpha}
\right)
\nonumber\\
&\qquad\qquad
-
P
\left(
t_{\vec r\alpha}^\dagger
t_{\vec r^{\,\prime}\alpha}
+
t_{\vec r^{\,\prime}\alpha}^\dagger
t_{\vec r\alpha}
\right)
+
P^2-Q^2
\Big].
\label{eq:mf_ham_q}
\end{align}

Using translational invariance and Fourier transformation, the Hamiltonian can be expressed in momentum space as

\begin{equation}
H
=
E_0
+
\sum_{\vec k}
\left[
\Lambda_{\vec k}
t_{\vec k\alpha}^\dagger
t_{\vec k\alpha}
+
\Delta_{\vec k}
\left(
t_{\vec k\alpha}^\dagger
t_{-\vec k\alpha}^\dagger
+
t_{\vec k\alpha}
t_{-\vec k\alpha}
\right)
\right],
\label{eq:hamiltonian_quad_appendix}
\end{equation}

where

\begin{align}
E_0 =&
\left(
-\frac{3}{4}J_d\bar{s}^2
-\mu \bar{s}^2
+\mu
\right)N_d
\nonumber\\
&
-
\frac{N_d}{4}
\left(
\frac{3J_1}{2}
+
J_2
+
J_3
\right)
(P^2-Q^2),
\end{align}

\begin{equation}
\Lambda_{\vec k}
=
\left(
\frac{J_d}{4}-\mu
\right)
+
2\xi_{\vec k},
\end{equation}

\begin{equation}
\Delta_{\vec k}
=
\frac{\bar{s}^2}{4}\xi_{\vec k}
-
\frac{Q}{4}\phi_{\vec k},
\end{equation}

with

\begin{equation}
\xi_{\vec k}
=
(-J_1+J_2)\cos k_y
+
\left(
-\frac{J_1}{2}+J_3
\right)
\cos 2k_x,
\end{equation}

and

\begin{equation}
\phi_{\vec k}
=
(J_1+J_2)\cos k_y
+
\left(
\frac{J_1}{2}+J_3
\right)
\cos 2k_x.
\end{equation}

The Hamiltonian is diagonalized using the Bogoliubov transformation discussed in the main text, leading to the quasiparticle dispersion

\begin{equation}
\omega_{\vec k}
=
\sqrt{
\Lambda_{\vec k}^2
-
4\Delta_{\vec k}^2
}.
\end{equation}

Minimization of the ground-state energy with respect to the mean-field parameters yields the self-consistent equations

\begin{equation}
\bar{s}^2
=
\frac{5}{2}
-
\frac{3}{2N_d}
\sum_{\vec k}
\frac{\Lambda_{\vec k}}{\omega_{\vec k}},
\end{equation}

\begin{equation}
\mu
=
-\frac{3}{4}J_d
+
\frac{3}{4N_d}
\sum_{\vec k}
\frac{
(\Lambda_{\vec k}-2\Delta_{\vec k})
\xi_{\vec k}
}{
\omega_{\vec k}
},
\end{equation}

\begin{equation}
P
=
\left[
\frac{3}{4N_d}
\sum_{\vec k}
\frac{
\Lambda_{\vec k}\phi_{\vec k}
}{
2\omega_{\vec k}
}
\right]
\frac{1}{
\frac{3J_1}{2}+J_2+J_3
},
\end{equation}

and

\begin{equation}
Q
=
\left[
-\frac{3}{4N_d}
\sum_{\vec k}
\frac{
\Delta_{\vec k}\phi_{\vec k}
}{
\omega_{\vec k}
}
\right]
\frac{1}{
\frac{3J_1}{2}+J_2+J_3
}.
\end{equation}

The effect of quartic triplet interactions on the triplon gap is shown in Fig.~\ref{fig:trip_int}. It is evident that the inclusion of triplet--triplet interactions significantly modifies the phase boundary associated with the transition from the dimerized phase to the Néel antiferromagnetic phase. Within the harmonic approximation, the triplon gap closes at $J_1 \approx 2.82$. However, upon incorporating the quartic interaction terms through the mean-field decoupling scheme, the gap closes at $J_1 \approx 2.36$. This substantial renormalization of the phase boundary on the antiferromagnetic side brings the bond-operator results into much closer agreement with the DMRG estimate of the critical point, $J_1 \approx 2.29$, obtained in the main text.

In contrast, the phase boundary between the double-stripe ordered phase and the dimerized phase remains nearly unaffected by the quartic corrections. As shown in Fig.~\ref{fig:trip_int}, the gap closing point changes only slightly, from $J_1 \approx -0.81$ within the harmonic approximation to $J_1 \approx -0.80$ after including the triplet interactions. This indicates that the effect of triplet--triplet interactions is considerably weaker for stripe--dimer transition.

\section{Finite-Size Scaling of the Magnetic Order Parameter}
\label{appendix:finite_size}

\begin{figure}
    \centering
    \includegraphics[width=0.88\columnwidth]{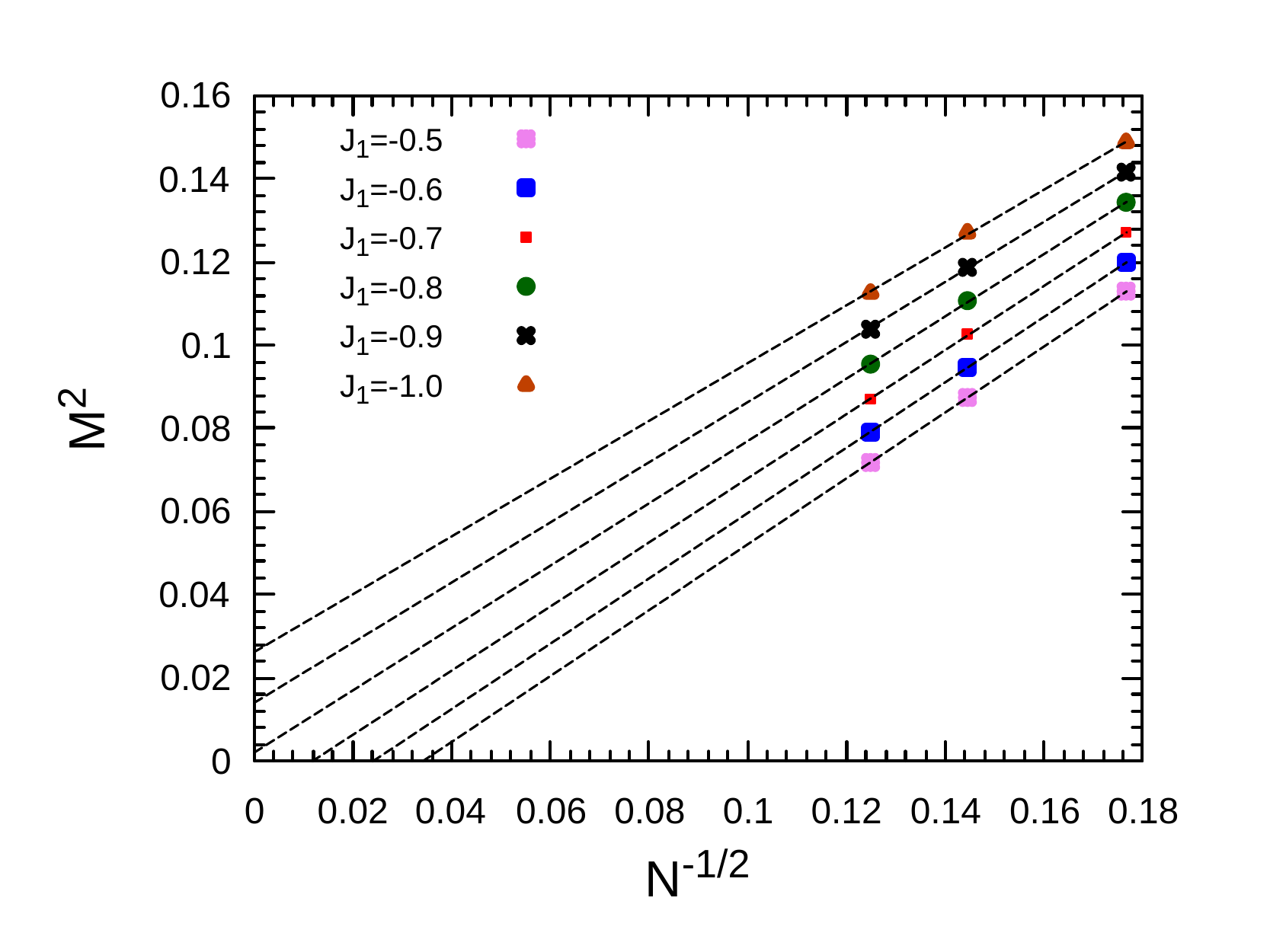}
    \caption{Finite-size scaling of the double-stripe magnetic order parameter for different values of $J_1$. The dashed lines represent linear extrapolations in $1/\sqrt{N}$. The extrapolated thermodynamic value decreases with increasing $J_1$ and vanishes near $J_1\approx -0.79$, indicating the disappearance of long-range double-stripe order.}
    \label{fig:finite_1}
\end{figure}

\begin{figure*}
    \centering
    \includegraphics[width=0.48\textwidth]{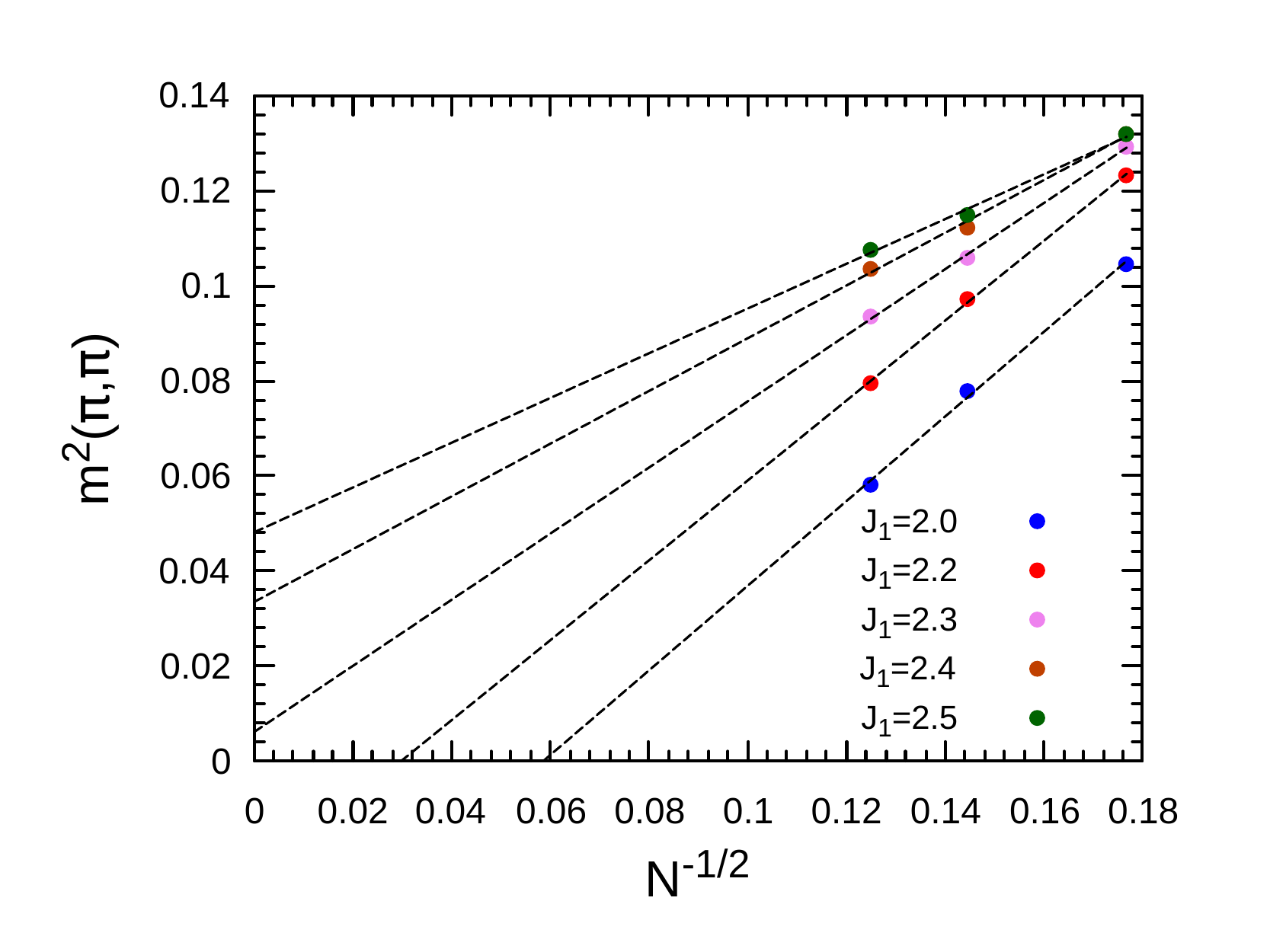}
    \includegraphics[width=0.48\textwidth]{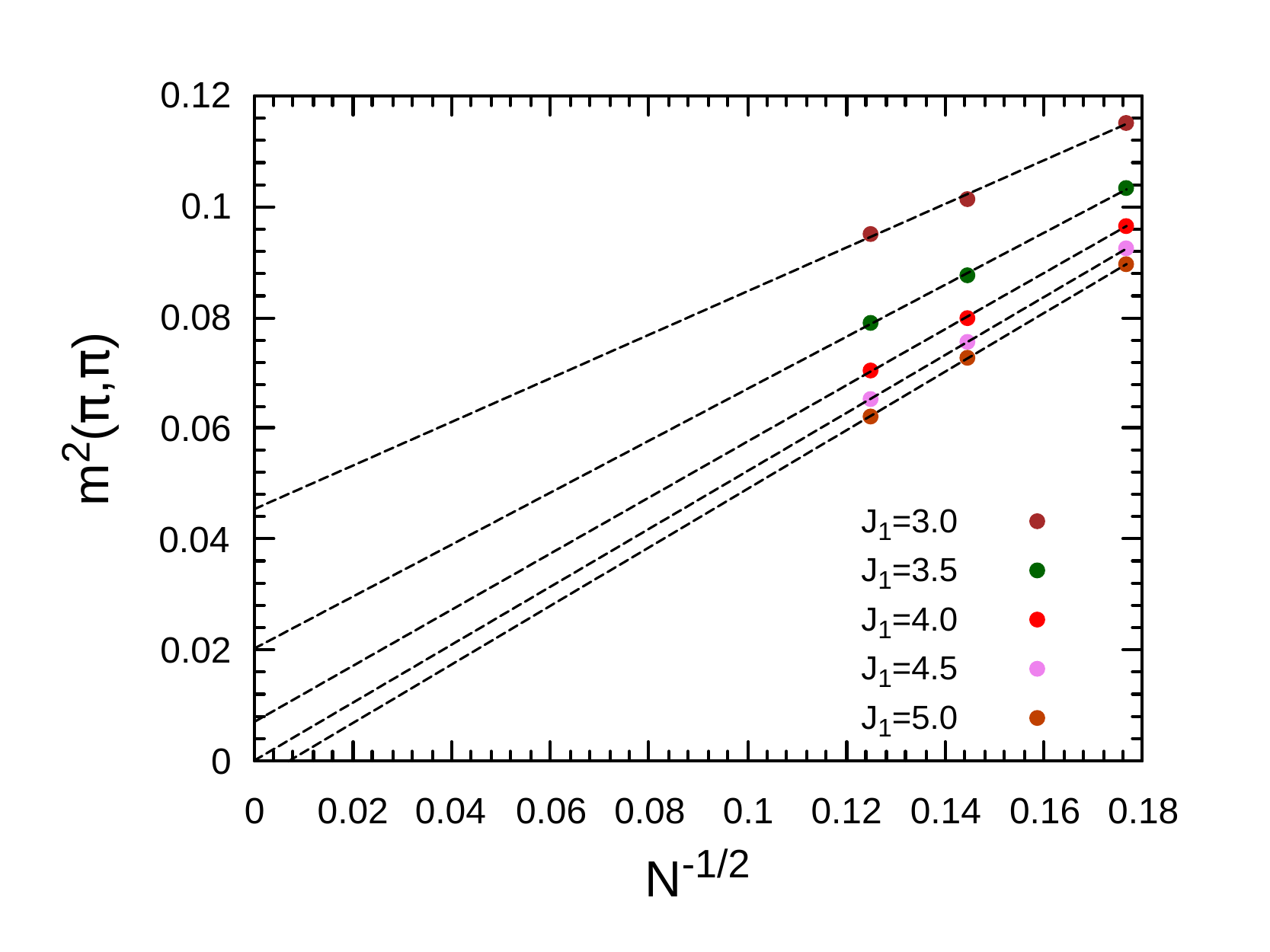}
    \caption{Finite-size scaling of the N\'eel magnetic order parameter obtained from linear extrapolation in $1/\sqrt{N}$. The extrapolated thermodynamic value approaches zero near the phase boundaries, indicating the suppression of N\'eel long-range order. The two panels correspond to different parameter regimes with critical points around $J_1\approx2.29$ and $J_1\approx4.5$, respectively.}
    \label{fig:tfinite_2}
\end{figure*}

To characterize the presence of long-range magnetic order in the thermodynamic limit, we perform a finite-size scaling analysis of the magnetic order parameter. In systems with spontaneous breaking of continuous spin rotational symmetry, the low-energy excitations are governed by Goldstone modes, which give rise to characteristic finite-size corrections to the order parameter. Following the general arguments based on the nonlinear sigma model description of ordered antiferromagnets~\cite{Schulz_1996,PhysRevB.39.2608}, the magnetic structure factor for a finite system of $N$ sites is expected to scale as

\begin{equation}
M_N^2(\mathbf{Q}) = M_0^2(\mathbf{Q}) + \frac{a}{\sqrt{N}} + \mathcal{O}\left(\frac{1}{N}\right),
\label{eq:fss_general}
\end{equation}

where $M_N^2(\mathbf{Q})$ denotes the magnetic structure factor at ordering wavevector $\mathbf{Q}$ for a system of size $N$, and $M_0^2(\mathbf{Q})$ is its value in the thermodynamic limit. The coefficient $a$ is a non-universal constant that depends on microscopic details and possible anisotropies of the ordered phase.

Equation~(\ref{eq:fss_general}) implies that the leading finite-size correction varies linearly with $1/\sqrt{N}$. Therefore, the thermodynamic value $M_0^2(\mathbf{Q})$ can be obtained by extrapolating the numerical data as a function of $1/\sqrt{N}$.

In the present work, we carried out this extrapolation using DMRG results obtained for clusters of sizes $8\times4$, $8\times6$, and $8\times8$. A linear fit of $M_N^2(\mathbf{Q})$ versus $1/\sqrt{N}$ was performed, and the intercept at $1/\sqrt{N}\rightarrow0$ provides an estimate of the magnetic order parameter in the thermodynamic limit.

Figure~\ref{fig:finite_1} shows the finite-size scaling of the double-stripe magnetic order parameter for several values of $J_1$. For sufficiently large ferromagnetic coupling, the extrapolated intercept remains finite and positive, indicating stable long-range double-stripe magnetic order in the thermodynamic limit. As $J_1$ increases toward weaker ferromagnetic values, the extrapolated intercept gradually decreases and eventually becomes negative near $J_1 \approx -0.79$. Since a negative value of the squared order parameter is unphysical, this indicates the disappearance of long-range double-stripe order and signals a transition out of the ordered phase.

Similarly, Fig.~\ref{fig:tfinite_2} presents the finite-size scaling analysis of the N\'eel order parameter. The extrapolated thermodynamic value decreases continuously upon approaching the phase boundary and vanishes near the critical coupling. The two panels correspond to different parameter regimes, where the extrapolated order parameter approaches zero around $J_1 \approx 2.29$ and $J_1 \approx 4.5$, respectively. The vanishing of the extrapolated intercept indicates the suppression of N\'eel long-range order and provides an estimate for the phase transition points.

We note that the precise coefficient of the finite-size correction depends on the symmetry properties of the ordered state and the associated low-energy field theory. However, the characteristic $1/\sqrt{N}$ dependence is expected to be generic for two-dimensional magnetically ordered phases with gapless Goldstone excitations.

\bibliographystyle{apsrev4-2}
\bibliography{ladder}
\end{document}